\documentclass[a4paper]{JHEP3}

\usepackage{graphicx}
\usepackage{amsmath}
\usepackage{cite}
\usepackage{axodraw}

\def\eqref#1{(\ref{#1})}
\def\text{\rm }
\def\Res{{\rm Res}}

\newcommand{\fnsz}{\footnotesize}
\newcommand{\beq}{\begin{equation}}
\newcommand{\eeq}{\end{equation}}
\newcommand{\bqa}{\begin{eqnarray}}
\newcommand{\eqa}{\end{eqnarray}}

\def\db#1{ D_{#1}}

\def\spa#1.#2{\langle#1\,#2\rangle}
\def\spb#1.#2{[#1\,#2]}

\def\spab#1.#2.#3{\langle\mskip-1mu{#1}
                  | #2 | {#3}]}

\def\spba#1.#2.#3{[\mskip-1mu{#1}
                  | #2 | {#3}\rangle}

\def\spbb#1.#2.#3.#4{[\mskip-1mu{#1}
                     | {#2} \ {#3} | {#4}]}

\def\spaa#1.#2.#3.#4{\langle\mskip-1mu{#1}
                     | {#2} \ {#3} | {#4}\rangle}

\newcommand{\bea}{\begin{eqnarray}}

\newcommand{\eea}{\end{eqnarray}}

\newcommand{\bean}{\begin{eqnarray*}}

\newcommand{\eean}{\end{eqnarray*}}
           
\newcommand{\nn}{\nonumber \\}

\title{On the Integrand-Reduction Method for Two-Loop Scattering Amplitudes}

\author{Pierpaolo Mastrolia \\
Max-Planck Insitut f\"ur Physik, F\"ohringer Ring, 6, D-80805 M\"unchen, Germany \\
Departamento de F\'isica Te\'orica, Universidad Aut\'onoma de Madrid, Cantoblanco, 
28049 Madrid, Spain \\
E-mail: \email{ppaolo@mppmu.mpg.de}
}    

\author{Giovanni Ossola \\ 
Physics Department, New York City College of Technology,\\ 
                      City University Of New York, 
           300 Jay Street, Brooklyn NY 11201, USA.\\
The Graduate School and University Center, The City University of New York
365 Fifth Avenue, New York NY 10016, USA \\
E-mail: \email{GOssola@citytech.cuny.edu}
}

\abstract{We propose a first implementation of the integrand-reduction method 
for two-loop scattering amplitudes.
We show that the residues of the amplitudes on multi-particle cuts 
are polynomials in the irreducible scalar products involving the loop momenta,
and that the reduction of the amplitudes in terms of master integrals can 
be realized through polynomial fitting of the integrand, 
without any apriori knowledge of the integral basis.
We discuss how the polynomial shapes of the residues determine the basis
of master integrals appearing in the final result.
We present a four-dimensional constructive algorithm 
that we apply to planar and non-planar contributions 
to the 4- and 5-point MHV amplitudes in ${\cal N}=4$ SYM.
The technique hereby discussed extends the well-established analogous 
method holding for one-loop amplitudes,
and can be considered a preliminary study towards the systematic 
reduction  at the integrand-level of two-loop amplitudes in any gauge theory,
suitable for their automated semianalytic evaluation.
}

\begin{document}

\section{Introduction}

Scattering amplitudes constitute the core of the perturbative structure of 
quantum field theories. The recent development of novel methods for computing them
has been highly stimulated by a deeper understanding 
of the multi-channel factorization properties naturally emerging as 
{\it reactions} of the amplitudes under deformations of the kinematics 
in the complex plane dictated by on-shell 
\cite{Cachazo:2004kj,Britto:2004ap} and 
generalized unitarity-cut conditions \cite{Bern:1994zx,Britto:2004nc}.

Analyticity and unitarity of scattering amplitudes \cite{OldUnitarity} have then been strengthened  
by the complementary classification of the mathematical structures 
present in the residues at the singularities, better understood 
after uncovering 
a quadratic recurrence relation for tree-level amplitudes, 
the so called BCFW-recursion \cite{Britto:2004ap}, 
its link to the leading singularity of one-loop amplitudes \cite{Britto:2004nc},
and a relation between 
numerator and denominators 
of one-loop Feynman integr{\it als}, yielding the multipole decomposition of 
Feynman integr{\it ands}, stronghold of the by-now known as OPP method
\cite{Ossola:2006us}.

These new insights, which stem from a reinterpretation 
of tree-level scattering within the twistor string theory \cite{Witten:2003nn},
have catalyzed the study of novel mathematical frameworks in the more 
supersymmetric sectors of quantum field 
theories, such as dual conformal symmetries 
\cite{Drummond:2006rz}, 
grassmanians 
\cite{ArkaniHamed:2009dn}, 
Wilson-loops/gluon-amplitudes duality 
\cite{Alday:2007hr}, 
color/kinematic 
and gravity/gauge dualities  \cite{Bern:2008qj,Bern:2010ue},
as well as on-shell \cite{Cachazo:2004kj,Britto:2004ap,Badger:2005zh,Bern:2005cq} 
and generalised unitarity-based methods 
\cite{Bern:1994zx,Britto:2004nc,GenUn,Ellis:2007br},
and more generally the breakthrough advances in automating 
the evaluation of multi-particle scattering one-loop amplitudes, as demanded by the
experimental programmes at hadron colliders.

In general, when a direct integration of Feynman integrals is prohibitive, 
the evaluation of scattering amplitudes beyond 
the leading order is addressed in two stages: 
{\it i)} the reduction in terms of an integral basis, and {\it ii)} the evaluation 
of the elements of such a basis, called {\it master integrals} (MI's).

At one-loop, 
the advantage of knowing {\it apriori} that the basis of MI's is formed by scalar 
one-loop functions \cite{Passarino:1978jh},
as well as the availability of their analytic expression \cite{vanOldenborgh:1989wn},
allowed the community to focus on the development of efficient algorithms  
for extracting the coefficients multiplying each MI's. 
Improved tensor decomposition \cite{Denner:2005nn}, 
complex integration and contour deformation \cite{Nagy:2006xy},
on-shell and generalised unitarity-based methods, and 
integrand-reduction techniques \cite{Ossola:2006us,Ellis:2007br,Mastrolia:2010nb,Heinrich:2010ax} 
led to results which only few years ago were considered inconceivable,
and to such a high level of automation \cite{Mastrolia:2010nb,Ossola:2007ax} 
that different scattering processes at the 
next-to-leading order accuracy can be handled by single, yet multipurpose, codes 
\cite{Hahn:2010zi,vanHameren:2009dr,Bevilacqua:2010mx,Hirschi:2011pa,GoSam}.

At higher-loop, and in particular at two-loop to begin with, the situation is different.
The basis of MI's is not known apriori. 
MI's are identified at the end of the reduction procedure, and afterwards the problem of their 
evaluation arises.
The most used multi-loop reduction technique is the well-known Laporta algorithm \cite{Laporta:2001dd},
based on the solution of algebraic systems of equations obtained through integration-by-parts 
identities \cite{Tkachov:1981wb}. 
The recent progress in evaluating amplitudes beyond one-loop 
has been necessarily accompanied by the improvement of mathematical methods dedicated to Feynman 
integrals, such as 
difference \cite{Laporta:2001dd,Lee:2010ug} 
and 
differential \cite{Kotikov:1991pm} 
equations, 
Mellin-Barnes integration \cite{Smirnov:1999gc}, asymptotic expansions \cite{Beneke:1997zp}, sector decomposition
\cite{Binoth:2000ps}, complex integration and contour deformation 
\cite{Anastasiou:2007qb} -- to list few of them. \\ 

In this paper we aim at extending the combined use of 
{\it unitarity-based methods} and 
{\it integrand-reduction},
in order to accomplish 
the semianalytic reduction of two-loop amplitudes to MI's.
The use of unitarity-cuts and complex momenta for on-shell internal particles 
turned unitarity-based methods into very efficient
tools for computing scattering amplitudes. These methods exploit two general properties of 
scattering amplitudes, such as analyticity and unitarity:
the former granting that amplitudes can be reconstructed from 
the knowledge of their (generalised) singularity-structure; the latter granting that 
the residues at the 
singular points factorize into products of simpler amplitudes.
Unitarity-based methods are founded on the underlying representation 
of scattering amplitudes as a linear combination of MI's, and their principle is 
the extraction of the coefficients entering in such a linear combination 
by matching the cuts of the amplitudes onto the cuts of each MI.

In the past years, general criteria to determine the MI's of arbitrary problems
have been investigated \cite{whoisMI}, and 
very recently, a minimal basis for two-loop planar integrals has been 
identified \cite{Gluza:2010ws}, through an ameliorated solution of systems 
of equations involving integration-by-parts identities and 
supplementary Gram-determinant relations.

Cutting rules as computational tools have been introduced at two-loop in the context 
of supersymmetric amplitudes \cite{Bern:1997nh} and later applied to the case of pure 
QCD amplitudes \cite{Bern:2000dn}.
The use of complex momenta for propagating particles to fulfill the multiple 
cuts of two-loop amplitudes has been proposed for extending the benefits of 
the one-loop quadruple-cut technique, 
to the octa-cut \cite{Buchbinder:2005wp} and the leading singularity 
techniques \cite{Cachazo:2008vp}, 
as well as the method of maximal cuts \cite{maximalcut}, all 
indicating the possibility of 
{\it ``reducing the computation of multi-loop amplitudes to the computation of residues 
(which end up being related to tree-amplitudes) 
and to the solution of linear systems''} \cite{Cachazo:2008vp} -- an idea we elaborate on hereby. \\

The multi-particle pole decomposition for the integrands 
of arbitrary scattering amplitudes emerges
from the combination of analyticity and unitarity 
with the idea of a reduction under the integral sign. 

The principle of an integrand-reduction method is 
the underlying multi-particle pole expansion for the integrand of any scattering amplitude,
or, equivalently, the relation between numerator and denominators of the integrand:
a representation where the numerator of each Feynman integral is expressed as a combination 
of products of the corresponding denominators, with polynomial coefficients. 

The key element in the integrand-decomposition is represented by 
the shape of the residues on the multi-particle pole before integration:
each residue is a (multivariate) polynomial in the 
{\it irreducible scalar products} (ISP's) formed by the loop momenta and 
either external momenta or polarization vectors constructed out of them; 
the scalar products appearing in the residues are 
by definition {\it irreducible}, namely 
they cannot be expressed in terms of the denominators of the integrand -- otherwise 
mutual simplifications may occur and the notion of residues to a specific set of vanishing 
denominators would become meaningless.

The polynomial structure of the multi-particle residues is a {\it qualitative} information
that turns into a {\it quantitative} algorithm for decomposing arbitrary 
amplitudes in terms of MI's at the integrand level. 
In the context of an integrand-reduction, any explicit integration procedure 
and/or any matching procedure between cuts of amplitudes and cuts of MI's 
is replaced by {\it polynomial fitting}, which is a simpler operation.

Decomposing the amplitudes in terms of MI's amounts to reconstructing the full polynomiality 
of the residues, {\it i.e.} it amounts to determining all the coefficients of each polynomial. \\

The main goal of this paper is to outline guiding criteria to constrain the 
polynomial form of the residue on each multiple-cut of an arbitrary two-loop amplitude.
Unlike the one-loop case, where the residues of the multiple-cut have been 
systematized for all the cuts, in the two-loop case, their form 
is still unknown.
Their existence is a prerequisite for establishing a relation between numerator 
and denominators of any two-loop integrand.
Their implicit form can be given in terms of unknown coefficients, which are 
determined through polynomial fitting.
As in the one-loop case,
the full reconstruction of the polynomial residues is engineered 
via a projection technique based on the Discrete Fourier Transform \cite{Mastrolia:2008jb}, 
and requires only the knowledge of the numerator evaluated at explicit values of 
the loop momenta as many times as the number of the unknown coefficients.

Another feature of the integrand-reduction algorithm we are describing is 
that the determination of the polynomial form of 
the residues amounts to choose a basis of MI's, which 
does not necessarily need to be known apriori.
In fact, as we will see, each ISP appearing in the polynomial residues
is the numerator of a potential MI which may appear in the final result
(other than the scalar integrals). We remark that 
the set of MI's which will emerge at the end of the integrand-reduction (as well as after applying 
any unitarity-based methods) is not necessarily the minimal set of basic integrals.  
Integration-by-parts identities, 
Lorentz-invariance identities, as well as 
Gram-determinant identities  may not be detected in the framework of a cut-construction, and 
therefore constitute additional, independent relations which can further reduce the number of MI's
which have to be evaluated after the reduction stage.

We define the $m-$fold cut of a diagram as the set of on-shell conditions corresponding to 
the vanishing of $m$ denominators present in that diagram. 
As in the maximal-cut method \cite{maximalcut}, we do not cut additional denominators 
which might arise from cutting one-loop sub-diagram.\\
We outline the driving principles for the cut-construction of the residues at the $m-$fold cut,
and the use of self-consistency checks that ensure the correctness of the reconstructed polynomials,
namely after the determination of their unknown coefficients.
These checks, called {\it local} and {\it global} $(N=N)$-tests, are analogous to the tests 
employed in the one-loop integrand-reduction \cite{Ossola:2007ax,Mastrolia:2010nb}, 
and monitor, respectively, the completeness 
of the polynomial residues, and the correctness of the final decomposition formula. \\
The values of the loop momenta used for the numerator sampling are chosen among the 
solutions of the corresponding $m-$fold cut.

We verify the integrand-reduction algorithm by applying its procedures to 
planar and non-planar contributions 
to the 4-point MHV \cite{Bern:1997nh} and 5-point MHV \cite{Bern:2006vw} 
amplitudes in ${\cal N}=4$ SYM, and derive an expression for the non-planar 
pentacross diagram in terms of master integrals.

This work can be considered as a first building block of a new technique, that 
once developed {\it in toto}, would allow for the semianalytic 
reduction of multi-loop amplitudes.

\section{Four-dimensional Reduction Algorithm}
\label{sec:Reduction}

The reduction method hereby presented extends to two-loop 
the {\it integrand-reduction} procedures originally elaborated 
for arbitrary one-loop scattering amplitudes
\cite{Ossola:2006us,Ellis:2007br}.

An arbitrary two-loop $n$-point amplitude in the dimensional regularization scheme 
can be written as
\bea
&& {\cal A}_n = 
\int d^{4-2\epsilon}q \int d^{4-2\epsilon} k \ 
A( q,  k) \ , \nn 
&& A(q, k)= 
\frac{N(q, k)
}{
\db{1}\db{1}\cdots\db{n}
} \ , \\
&& \db{i} = (\alpha_i q + \beta_i k + p_i)^2-m_i^2 , 
\qquad \alpha_i,\beta_i \in \{0,1\} 
\nonumber
\label{def:An}
\eea
where $\epsilon = (4-d)/2$, and $d$ is the continuous-dimensional parameter.
Extra-dimensional components of the loop momenta can be parametrized as pseudo-mass
variables, $\lambda_q$ and $\lambda_k$, one for each loop momenta, according 
to the following scheme,
\bea
q\!\!\!\slash_{4-2\epsilon} &=& q\!\!\!\slash_4 + i \lambda_q \gamma_5 \ , 
\quad q^2_{4-2\epsilon} = q_4^2 - \lambda_q^2 \ , \\
k\!\!\!\slash_{4-2\epsilon} &=& k\!\!\!\slash_4 + i \lambda_k \gamma_5 \ ,
\quad k^2_{4-2\epsilon} = k_4^2 - \lambda_k^2 \ , \\
\int d^{4-2\epsilon}q \int d^{4-2\epsilon} k &=&
\int d^{-2\epsilon} \lambda_q \ 
\int d^{-2\epsilon} \lambda_k \ 
\int d^{4} q_4 \ \int d^{4} k_4 \ .
\eea
As in the one-loop case, $\lambda_q^2$ and $\lambda_k^2$ 
would appear as additional variables in the polynomial residues, and, consequently,
would be responsible for the appearance of MI's in higher dimensions \cite{Bern:2000dn}. 
The following discussion is limited to a purely four-dimensional reduction,
in which the loop variables $q_\mu$ and $k_\mu$ are defined in four dimensions.

\subsection{Integrand Decomposition}

The stronghold of the integrand-reduction of an arbitrary two-loop $n$-point amplitude
is the decomposition of the numerator $N(q,k)$
in terms of denominators $\db{i}$ for $i=1,\ldots,n$.
Following the same pattern as in the one-loop case, 
a plausible {\it ansatz} reads,
\bea
\label{eq:2}
N({q,k}) &=&
\sum_{i_1 < \!< i_8}^{n}
          \Delta_{i_1, \ldots, i_8}({q,k})
\prod_{h \ne i_1, \ldots, i_8}^{n} \db{h} 
+\sum_{i_1 < \!< i_7}^{n}
          \Delta_{i_1, \ldots, i_7}({q,k})
\prod_{h \ne i_1, \ldots, i_7}^{n} \db{h} 
+ \nn     &+&
\ldots 
+\sum_{i_1 < \!< i_2}^{n}
          \Delta_{i_1, i_2}({q,k}) 
\prod_{h \ne i_1 , i_2}^{n} \db{h} \ , \qquad
\label{def:MOT:deco}
\eea
where $ i_1 < \!< i_8 $ stands for a lexicographic ordering 
$i_1 < i_2 < \ldots < i_7 < i_8$, and where the $\Delta$'s are functions 
depending on the loop momenta.
By using the decomposition (\ref{def:MOT:deco}) in Eq.(\ref{def:An}), 
the multi-pole nature of the integrand of an arbitrary two-loop $n$-point 
amplitude becomes manifest,
\bea
A(q,k) &=&
\sum_{i_1 < \!< i_8}^{n}
         { \Delta_{i_1, \dots, i_8}({q,k}) \over 
           \db{i_1} \db{i_2} \ldots \db{i_8} } 
+
\sum_{i_1 < \!< i_7}^{n}
         { \Delta_{i_1, \dots, i_7 }({q,k}) \over 
           \db{i_1} \db{i_2} \ldots \db{i_7} } 
+
\ldots 
+
\sum_{i_1 < \!< i_2}^{n}
         { \Delta_{i_1, i_2}({q,k})  \over
           \db{i_1} \db{i_2} } \ . \qquad
\label{def:MOT:ampdeco}
\eea
The above expression, upon integration, yields the decomposition of the amplitude
in terms of Master Integrals (MI's), respectively associated to diagrams 
with 8-, 7-, \ldots, 2-denominators,
namely down to the products of two 1-point functions (one tadpole for each loop).
In Eq.(\ref{def:MOT:ampdeco}) each function 
$\Delta(q,k)$ parametrizes the residue of the amplitude on the multi-particle cut
that corresponds to the set of vanishing denominators it is sitting on.
The four-dimensional decomposition in 
Eqs.(\ref{def:MOT:deco},\ref{def:MOT:ampdeco}) begins with 
8-denominator terms. They correspond to the {\it maximal singularities} of two-loop amplitudes
in four dimensions, accessed by freezing both integration momenta with 
the simultaneous vanishing of eight denominators.
We expect that for dimensionally regulated amplitudes, 
due to the presence of additional degrees of freedom, 
$\lambda_q$ and $\lambda_k$,
an extended integrand-decomposition formula should hold, 
where higher-denominator functions are accommodated. 
Extension of the presented method to two-loop amplitudes in dimensional regularization 
will be the subject of a future work.

\subsection{Residues}
\label{sec:Polynomials}

We define the $m-$fold cut of a diagram
as the set of on-shell conditions corresponding to 
the vanishing of $m$ denominators present in that diagram.
The calculation of a generic scattering amplitude amounts 
to the problem of extracting 
the coefficients of multivariate polynomials, 
generated at every step of the multiple-cut analysis.
In fact, we will see that each $\Delta(q,k)$ is polynomial 
in the scalar products of the loop momenta with either external momenta  
or polarization vectors constructed out of them. These scalar products 
cannot be expressed in terms of the denominators $D_i$, and therefore 
are defined {\it irreducible scalar products} (ISP's).
In the case of the residue to an $m-$fold cut, $\Delta_{i_1, \dots, i_m}$, with $m<8$,
the ISP's correspond to the components of the loop momenta not frozen 
by the on-shell conditions; the eightfold-cut conditions of a two-loop amplitude 
freeze completely both integration momenta, like the quadruple-cut in the one-loop case.

The polynomial form of $\Delta_{i_1, \dots, i_m}$ depends on the number of 
independent {\it external} momenta of the $n$-point diagram 
identified by $\db{i_1} \ldots \db{i_m}$. 
In an $n$-point diagram, due to momentum conservation, only $(n-1)$ external
momenta are independent.
In four dimensions, there can be at most four independent external momenta.
Therefore, despite the number of loops, we can trivially cast  
$n-$point amplitudes in two groups according to whether $n$ is larger than 4 or not:
%
% the 10 commandments
%
\begin{itemize}
\item In the case of $n-$point diagrams with $n \ge 5$, four (out of $n$) external momenta 
can be chosen to form a {\it real} four-dimensional vector basis.

\item In the case of $n-$point diagrams with $n \le 4$, 
we can choose only up to three (out of $n$) independent external momenta.
But they are not sufficient, and additional elements, {\it orthogonal} to them, 
have to be taken into account to {\it complete} the four dimensional basis.
In this case, one can use complex polarization vectors as orthogonal complement, 
to form a {\it complex} basis.

\item Each $m-$fold cut will be characterized by one of these two kinds of basis,
and the loop momenta will be decomposed along the vectors forming it.

\item The residue of an $m-$fold cut, $\Delta_{i_1, \dots, i_m}$, is polynomial in 
the components of the loop momenta, and therefore it is polynomial in the 
ISP's constructed from the loop momenta and the elements 
of the basis used to decompose them.

\item 
The polynomial form of the residue $\Delta_{i_1, \dots, i_m}$ 
determines the MI's potentially appearing in the final decomposition.
\end{itemize}

\subsubsection{Polynomial Structures and Master Integrals}

Let's consider $q$ and $k$ as the solutions of the $m$-fold cut identified by 
the vanishing of $\db{i_1}\ldots \db{i_m}$.
We can decompose the loop momentum $q$ in terms of the basis $\{\tau_i\}$ 
and $k$ along the basis $\{ e_i \}$,
\bea
 q^\mu &=& - p_0^\mu + x_1 \tau_1^\mu + x_2 \tau_2^\mu + x_3 \tau_3^\mu + x_4 \tau_4^\mu \ , \\ 
\label{eq:q_deco}
 k^\mu &=& - r_0^\mu + y_1 e_1^\mu + y_2 e_2^\mu + y_3 e_3^\mu + y_4 e_4^\mu \ . 
\label{eq:k_deco}
\eea 
where $p_0$ and $r_0$ are combinations of external momenta.

For simplicity, let us assume that $\Delta_{i_1, \dots, i_m}$ is  
a one-dimensional polynomial. The multivariate extension follows the same principles.
We can write $\Delta_{i_1, \dots, i_m}$ in terms of scalar products $(e_i \cdot (k+r_0))$ as
\bea
\Delta_{i_1, \dots, i_m}  = \sum_j c_j \ (e_i \cdot (k+r_0))^j \ .
\eea
Since $\Delta_{i_1, \dots, i_m}$ is the polynomial sitting on the denominators
$\db{i_1}\ldots \db{i_m}$,
the integral expression,
\bea
\int d^4q \int d^4k \ { \Delta_{i_1, \dots, i_m} \over \db{i_1}\ldots \db{i_m}} = 
\sum_j c_j \int d^4q \int d^4k \ {(e_i \cdot (k+r_0))^j \over \db{i_1}\ldots \db{i_m}} \ ,
\eea
can generate MI's. 
In general, there are two possibilities which can be encountered:
\bea
i) && \int d^4q \int d^4k \  {(e_i \cdot (k+r_0))^j \over \db{i_1}\ldots \db{i_m}} \ne 0 \ ; \\
ii) && \int d^4q \int d^4k \  {(e_i \cdot (k+r_0))^j \over \db{i_1}\ldots \db{i_m}} = 0 \ . 
\label{eq:mastercases}
\eea
In case $(i)$, the {\it natural} integral basis 
associated to the set of denominators $\db{i_1}\ldots \db{i_m}$ is determined
by the ISP $(e_i \cdot (k+r_0))$. According to the power $j$, we identify the following MI's:
\bea
j=0: &&\int d^4q \int d^4k \ {1 \over \db{i_1}\ldots \db{i_m}} \qquad {\rm (scalar \ MI)}\\
j=1: &&\int d^4q \int d^4k \ {(e_i \cdot (k+r_0)) \over \db{i_1}\ldots \db{i_m}} \qquad {\rm (linear \ MI)} \\
j=2: &&\int d^4q \int d^4k \ {(e_i \cdot (k+r_0))^2 \over \db{i_1}\ldots \db{i_m}} \qquad {\rm (quadratic \ MI)}\\
\ldots \quad  && \ldots \nonumber
\eea
Option $(ii)$ may occur in the case of an $m-$fold cut of an $n$-point diagram with 
$n \le 4$, where
the vector $e_i^\mu$ could be orthogonal to the independent vectors of the diagrams,
yielding the vanishing of the integral in Eq.(\ref{eq:mastercases}). 
Borrowing the terminology from the one-loop case, the ISP is {\it spurious} 
and does not generate any MI, other than the scalar one.

We observe that in the one-loop case the polynomial residues, from the single-cut 
up to the quadruple-cut, are written in terms of spurious ISP's. 
Only the pentuple-cut could have been written in terms
of non-spurious ISP's. However, at the one-loop level there is no room for non-spurious ISP, 
because any scalar product between the loop variables and external momenta is reducible. 
In fact, this is another reason for 
the one-loop 5-point amplitude in four dimensions to admit a complete reduction in terms of 
4-point integrals.

\subsubsection{Numerator Sampling}

As in the one-loop integrand-reduction,
once the polynomial form of the residues is given in terms of unknown coefficients, 
the numerator decomposition in Eq.(\ref{def:MOT:deco}) becomes the engine for 
determining them. \\
Since the function on the {\it l.h.s} of Eq.(\ref{def:MOT:deco}), namely the numerator $N(q,k)$
of an arbitrary two-loop diagram is a known quantity, and the {\it r.h.s} of 
Eq.(\ref{def:MOT:deco}) has a canonical representation in terms of $\Delta$-polynomials
and denominators $\db{i_1} \ldots \db{i_m}$, 
the determination of the unknown coefficients sitting in 
each of the $\Delta_{i_1, \dots, i_j}$ can be achieved by solving a system of linear equations.
This set of equations is obtained evaluating the {\it l.h.s} and the {\it r.h.s} of Eq.(\ref{def:MOT:deco})
for explicit values of the loop variables, as many times as the number of unknown coefficients
to be determined.
The numerator {\it sampling} can be performed either numerically or analytically. \\
Choosing the solutions 
of the $m-$fold cuts as sampling values is very convenient, because it allows a 
direct determination of the coefficients appearing in the corresponding 
residue $\Delta_{i_1, \dots, i_m}$. 
Also, the polynomial fitting can be better achieved by a projection technique based on 
the Discrete Fourier Transform \cite{Mastrolia:2008jb}, instead of inverting a system.

\subsubsection{Momentum Basis}

Once the polynomial form of each $\Delta_{i_1, \dots, i_m}$ is established, 
in terms of spurious or non-spurious ISP's,
one can use any basis to decompose the loop momenta $q$ and $k$ fulfilling the 
on-shell conditions $\db{i_1}=\ldots=\db{i_m}=0$. 
We find it convenient to use always a {\it complex} four dimensional basis 
\cite{Ossola:2006us},
which directly exposes the spurious character of ISP's, should they be present.

For each cut, we decompose the loop momenta $q$ and $k$, 
by means of two specific basis of four massless vectors.
We begin with the decomposition of the loop momentum $q^\mu$, 
flowing through an on-shell denominator carrying momentum $(q+p_0)$.
Let us construct the massless vectors $\tau_1$ and $\tau_2$ as a linear combination of 
the two external legs, say $K_1$ and $K_2$, 
\bea
 \tau_1^\mu = {1 \over \beta}\bigg(K_1^\mu + {K_1^2 \over \gamma} K_2^\mu \bigg) \ , \qquad
 \tau_2^\mu = {1 \over \beta}\bigg(K_2^\mu + {K_2^2 \over \gamma} K_1^\mu \bigg) \ ,  
\eea
with
\bea
\beta = 1 - {K_1^2 K_2^2 \over \gamma^2} \ , \quad {\rm and} \qquad
\gamma = K_1 \cdot K_2 
        + {\rm sgn}(1,K_1 \cdot K_2) \sqrt{ (K_1 \cdot K_2)^2 - K_1^2 K_2^2 } \ .
\eea
\noindent
Then, one builds the massless vectors $\tau_3$ and $\tau_4$ from $\tau_1$ and $\tau_2$
\bea
\tau_3^\mu = {\langle \tau_1| \gamma^\mu | \tau_2 ] \over 2} \ , \qquad 
\tau_4^\mu = {\langle \tau_2| \gamma^\mu | \tau_1 ] \over 2} \ ,
\eea
such that
\bea
 \tau_i^2 = 0 \ , \quad
 \tau_1 \cdot \tau_3 = \tau_1 \cdot \tau_4 = 0 \quad
 \tau_2 \cdot \tau_3 = \tau_2 \cdot \tau_4 = 0 \quad
 \tau_1 \cdot \tau_2 = - \tau_3 \cdot \tau_4 \ .
\eea
The basis $\tau_i$ can be used to decompose the loop-momentum $q$, as
\bea
 q^\mu = - p_0^\mu + x_1 \tau_1^\mu + x_2 \tau_2^\mu + x_3 \tau_3^\mu + x_4 \tau_4^\mu \ ,
\label{eq:q_deco}
\eea 

We double this procedure, by introducing a second basis $e_i$ to decompose the
other loop momentum, 
\bea
 k^\mu = - r_0^\mu + y_1 e_1^\mu + y_2 e_2^\mu + y_3 e_3^\mu + y_4 e_4^\mu \ .
\label{eq:k_deco}
\eea 
where the basis $e_i$ can be constructed through the same definitions
as $\tau_i$ by replacing $(K_1,K_2)$ with another couple of external vectors.

The two basis $\{\tau_i\}$ and $\{e_i\}$ are adopted for decomposing 
the solutions of the multiple-cuts of the two-loop amplitude.
In the following, the decompositions of $q$ and $k$ will be 
uniquely defined by specifying, case by case, $p_0$, $\tau_1$, $\tau_2$, and
$r_0$, $e_1$, $e_2$, respectively.

\subsection{Testing the Integrand-Decomposition}

The integrand-reduction algorithm offers 
self-consistency checks that ensure the correctness of the reconstructed polynomials 
after the determination of their unknown coefficients.
These checks, called {\it local} and {\it global} $(N=N)$-tests, are analogous to the tests 
employed in the one-loop integrand-reduction \cite{Ossola:2007ax,Mastrolia:2010nb}. 
We simply recall here their definitions:
\begin{itemize}
\item The {\it local $(N=N)$-test} monitors the completeness of the polynomial residues,
namely it can be used to verify that all the necessary ISP's, and their powers, have
been properly accounted for.
Accordingly, after the determination of the polynomial coefficients of a given residue,
$\Delta_{i_1, \dots, i_m}$, one can use any solutions of the corresponding $m$-fold cut-conditions 
other than the ones used for the determination of its coefficients, 
to verify the fulfillment of Eq.(\ref{def:MOT:deco}).
This test can be implemented {\it locally}, cut-by-cut,
and its failure means an incomplete parametrization of the residue.\\
We observe that there is only one case
where the {\it local $(N=N)$-test} cannot be applied: the maximal-cut of a given loop,
defined as the cut where all loop momenta are frozen by the on-shell cut-conditions.
In this case, in fact, the number of solutions of the maximal cut is finite,
and all of them might have been used for the determination of the polynomial coefficients.\\
Apart from that, the {\it local $(N=N)$-test} is a very powerful tool 
for the classification of the polynomial structures characterizing the residue of 
the $m-$fold cut for arbitrary amplitudes, at any loop, for any $m$ (other than maximal).

\item The {\it global $(N=N)$-test} ensures the correctness of the overall integrand-decomposition.
Accordingly, after the determination of all polynomial coefficients appearing in the {\it r.h.s.}
of Eq.(\ref{def:MOT:deco}), the identity between {\it l.h.s.} and {\it r.h.s.} of 
Eq.(\ref{def:MOT:deco}) must hold for arbitrary values of the loop variables. 
One can therefore verify it:
either {\it i)} at the end of the reduction procedure,
or {\it ii)} during the reduction, in order to rule out any further contribution 
possibly coming from sub-diagram MI's. In the latter case, the failure of the test indicates 
that other contributions are missing, and the reduction should continue for their detection.
\end{itemize}

\noindent
In the next sections we will present a series of examples that 
illustrate the integrand-reduction algorithm explicitly.
The required spinor-algebra has been implemented in {\it Mathematica}, using the package
{\it S@M} \cite{Maitre:2007jq}.

\section{Four-gluon MHV Amplitude in ${\cal N}=4$ SYM}

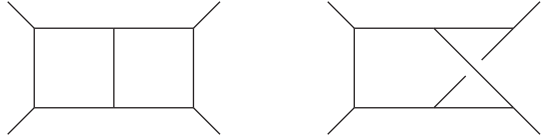
\begin{figure}[h]
\vspace*{1.5cm}
\begin{center}
%%%%%%%%%%%%%%%%%%%%%%%%%%%%%%%%%%%%%%%%
\begin{picture}(0,0)(0,0)
\SetScale{1.0}
\SetWidth{0.5}
\Line(0,15)(30,15)
\Line(30,15)(30,-15) %\Text(60,0)[]{{\fnsz{$k$}}}
\Line(30,-15)(0,-15)
\Line(0,15)(0,-15)
\Line(0,15)(-30,15)
\Line(-30,15)(-30,-15) %\Text(-60,0)[]{{\fnsz{$q$}}}
\Line(-30,-15)(0,-15)
\Line(30,15)(40,25)     %\Text(65,45)[]{{\fnsz{$1^-$}}}
\Line(30,-15)(40,-25)   %\Text(65,-45)[]{{\fnsz{$2^-$}}}
\Line(-30,15)(-40,25)   %  \Text(-65,45)[]{{\fnsz{$4^+$}}}
\Line(-30,-15)(-40,-25)  % \Text(-65,-45)[]{{\fnsz{$3^+$}}}
\end{picture}
%%%%%%%%%%%%%%%%%%%%%%%%%%%%%%%%%%%%%%%%
\hspace*{4cm}
%%%%%%%%%%%%%%%%%%%%%%%%%%%%%%%%%%%%%%%%
\begin{picture}(0,0)(0,0)
\SetScale{1.0}
\SetWidth{0.5}
\Line(0,15)(30,15)
\Line(0,15)(30,-15) %\Text(8,+8)[]{{\fnsz{$k$}}}
\Line(30,-15)(0,-15)
\Line(30,15)(18,3) 
%\CArc(15,0)(7,225,45) 
\Line(12,-3)(0,-15)
\Line(0,15)(-30,15)
\Line(-30,15)(-30,-15) %\Text(-60,0)[]{{\fnsz{$q$}}}
\Line(-30,-15)(0,-15)
\Line(30,15)(40,25)     %\Text(65,45)[]{{\fnsz{$1^-$}}}
\Line(30,-15)(40,-25)   %\Text(65,-45)[]{{\fnsz{$2^-$}}}
\Line(-30,15)(-40,25)    % \Text(-65,45)[]{{\fnsz{$4^+$}}}
\Line(-30,-15)(-40,-25)   %\Text(-65,-45)[]{{\fnsz{$3^+$}}}
\end{picture}
%%%%%%%%%%%%%%%%%%%%%%%%%%%%%%%%%%%%%%%%
\end{center}
\vspace*{1.0cm}
\caption{Two Master Integrals of the 4-gluon MHV amplitude in ${\cal N}=4$ SYM: 
the ladder (left) and crossed (right) $s$-channel double-box.}
\label{fig:4p:MIs}
\end{figure}

The 4-gluon MHV amplitude in ${\cal N}=4$ supersymmetric gauge theory 
was originally calculated in Ref.\cite{Bern:1997nh}. Two MI's 
appearing in the result are the planar and the crossed double-box, 
shown in Fig.\ref{fig:4p:MIs}.

\subsection{A contribution to the Ladder Amplitude}
\label{sec:4p:ladder}

\begin{figure}[t]
\vspace*{1.5cm}
\begin{center}
\hspace*{2.5cm}
\begin{minipage}{5cm}
%%%%%%%%%%%%%%%%%%%%%%%%%%%%%%%%%%%%%%%%
\begin{picture}(0,0)(0,0)
\SetScale{1.5}
\SetWidth{0.8}
\Line(0,15)(30,15)
\ArrowLine(30,15)(30,-15) \Text(60,0)[]{{\fnsz{$k$}}}
\Line(30,-15)(0,-15)
\Line(0,15)(0,-15)
\Line(0,15)(-30,15)
\ArrowLine(-30,15)(-30,-15) \Text(-60,0)[]{{\fnsz{$q$}}}
\Line(-30,-15)(0,-15)
\Line(30,15)(40,25)     \Text(65,45)[]{{\fnsz{$1^-$}}}
\Line(30,-15)(40,-25)   \Text(65,-45)[]{{\fnsz{$2^-$}}}
\Line(-30,15)(-40,25)     \Text(-65,45)[]{{\fnsz{$4^+$}}}
\Line(-30,-15)(-40,-25)   \Text(-65,-45)[]{{\fnsz{$3^+$}}}
\DashLine(15,20)(15,-20){2}
\DashLine(-15,20)(-15,-20){2}
\DashLine(-35,0)(35,0){2}
\Vertex(30,15){2}
\Vertex(30,-15){2}
\Vertex(-30,15){2}
\Vertex(-30,-15){2}
\Vertex(0,15){2}
\Vertex(0,-15){2}
\Text(35,27)[]{{\fnsz{$^+$}}}
\Text(10,27)[]{{\fnsz{$^-$}}}
\Text(-10,27)[]{{\fnsz{$^+$}}}
\Text(-35,27)[]{{\fnsz{$^-$}}}
\Text(35,-30)[]{{\fnsz{$^+$}}}
\Text(10,-30)[]{{\fnsz{$^-$}}}
\Text(-10,-30)[]{{\fnsz{$^+$}}}
\Text(-35,-30)[]{{\fnsz{$^-$}}}
\Text(52,15)[]{{\fnsz{$^+$}}}
\Text(52,-15)[]{{\fnsz{$^-$}}}
\Text(5,15)[]{{\fnsz{$^-$}}}
\Text(5,-15)[]{{\fnsz{$^+$}}}
\Text(-55,15)[]{{\fnsz{$^+$}}}
\Text(-55,-15)[]{{\fnsz{$^-$}}}
\end{picture}
%%%%%%%%%%%%%%%%%%%%%%%%%%%%%%%%%%%%%%%%
\end{minipage}
\begin{minipage}{3cm}
{\footnotesize
\bea
D_1 &=& k^2 \nn
D_2 &=& (k+p_2)^2 \nn
D_3 &=& (k-p_1)^2 \nn
D_4 &=& q^2 \nn
D_5 &=& (q+p_3)^2 \nn
D_6 &=& (q-p_4)^2 \nn
D_7 &=& (q+k+p_2+p_3)^2 \ . \nonumber
\eea
}
\end{minipage}
\end{center}
%\vspace*{1.0cm}
\caption{7fold-cut of the 4-point ladder diagram ($s$-channel)}
\label{fig:4p:ladder}
\end{figure}

Let us consider the contribution to the (leading-color) 
4-gluon MHV amplitude in ${\cal N}=4$ SYM coming from 
the helicity configuration depicted in Fig.\ref{fig:4p:ladder}, 
where only gluons circulate in both loops.
This case has been discussed in the context of generalised unitarity-based method 
in Ref. \cite{Buchbinder:2005wp}. 
Here we reproduce the same result, and 
discuss its decomposition in terms of the 
planar double-box MI in Fig.\ref{fig:4p:MIs} (left), 
achieved by integrand-reduction.

\paragraph{On-shell Solutions.}
The solutions of the 4-point 7fold-cut, $\db{1} = \ldots = \db{6} = \db{7} = 0$ 
can be decomposed according to 
Eqs.(\ref{eq:q_deco},\ref{eq:k_deco}), with the following definitions:
\bea
&& 
p_0^\mu = 0^\mu \ , \qquad
e_1^\mu = p_1^\mu \ , \qquad 
e_2^\mu = p_2^\mu \ , \qquad \\
\label{def:lad:ebasis7fold:1234567}
&& 
r_0^\mu = 0^\mu \ , \qquad
\tau_1^\mu = p_3^\mu \ , \qquad 
\tau_2^\mu = p_4^\mu\ . \qquad 
\label{def:lad:wbasis7fold:1234567}
\eea
In this case, the seven on-shell conditions, 
$\db{1} = \ldots = \db{6} = \db{7} = 0$, cannot freeze the loop momenta.
The parametric solution reads,
\bea
q_{(7)}^\mu = x_4 \tau_4^\mu \ , \qquad k_{(7)}^\mu = y_3 e_3^\mu \ ,
\eea
where one of the on-shell cut-conditions 
imposes a non-linear relation among $x_4$ and $y_3$, which can be implicitly written as 
$x_4 = x_4(y_3)$.
We choose $y_3$, namely the component of $k$ along $e_3$, as the variable
parametrizing the infinite set of solutions of the 7fold-cut.
%(q_i^{(7)},k_i^{(7)})$, with $i=1,\ldots,\infty$.

\paragraph{Residue.}
The residue of this 4-point 7fold-cut is defined as,
\bea
\Delta_{1234567}(q,k) = 
\Res_{1234567}\{
N({q,k})
\}  \ ,
\label{def:lad:Resi7:1234567:left}
\eea
where $\Res_{1234567}\{N({q,k})\}$ is the product of the six 
3-point tree-amplitudes sitting in the vertices exposed by the cuts.
The definition of each 3-point tree-amplitude can be derived from the general
expression,
\bea
A_3^{\rm tree}(1^-,2^-,3^+) = 
i {\spa{1}.{2}^3\over\spa{2}.{3} \spa{3}.{1}} 
\Bigg( {\spa{2}.{3} \over \spa{1}.{2}}\Bigg)^a
\label{def:3p:tree}
\eea
with $a=0,1,2$ respectively for gluons, fermions, and scalars. 
In the case at hand, $a=0$. 

To find the polynomial expression of $\Delta_{1234567}$, we use the following 
criteria.
\begin{enumerate}
\item{\it Diagram topology and Vector basis.}

In this 4-point double-box, only three external vectors are independent,
therefore any possible basis involving them 
needs to be completed by an additional orthogonal vector $\omega_3$, 
defined as,
\bea
\omega_3^\mu = 
- { (p_2 \cdot \tau_3) \tau_4^\mu - (p_2 \cdot \tau_4) \tau_3^\mu \over \tau_3 \cdot \tau_4 } 
= 
 { (p_3 \cdot e_3) e_4^\mu - (p_3 \cdot e_4) e_3^\mu \over e_3 \cdot e_4 } \ . 
\label{eq:def:omega3}
\eea

\item{\it Irreducible Scalar Products.} 

The ISP's which can be formed by the loop variables 
and the external momenta can be chosen to be
$(q\cdot p_2)$ and $(k \cdot p_3)$. 
Due to the explicit expression of the on-shell solutions, 
on the cut, one has 
\bea
(q\cdot p_2) &\to& 
(q_{(7)} \cdot p_2) 
= (q_{(7)} \cdot \omega_3)
= - x_4 (\tau_4 \cdot p_2) \ , \\
(k\cdot p_3) &\to& 
(k_{(7)} \cdot p_3) 
= (k_{(7)} \cdot \omega_3) 
= y_3 (e_3 \cdot p_3)  \ .
\label{eq:4p:ladder:ISPs}
\eea
\end{enumerate}

\noindent
Therefore, $\Delta_{1234567}$ is as a polynomial in terms
of $x_4$ and $y_3$, and can be written as
\bea
\Delta_{1234567}(q,k) = 
c_{1234567,0} 
+ \sum_{i=1}^{r} c_{1234567,i}   \ (\omega_3 \cdot q)^i 
+ \sum_{i=1}^{r} c_{1234567,i+r} \ (\omega_3 \cdot k)^i \ ,
\label{def:lad:Resi7:1234567:right}
\eea
where $r$ is the maximum rank in the integration momenta.
Due to the orthogonality between $\omega_3$ and the 
external momenta, we identify two classes 
of vanishing integrals,
\bea
\int d^4q \int d^4k {(\omega_3 \cdot q)^n \over \db{1} \ldots \db{7}} &=& 0 \ , \\
\int d^4q \int d^4k {(\omega_3 \cdot k)^n \over \db{1} \ldots \db{7}} &=& 0 \ .
\eea
As a result, the ISP's $(\omega_3 \cdot q)$ and $(\omega_3 \cdot k)$ 
are {\it spurious}, and the polynomial form in 
Eq.(\ref{def:lad:Resi7:1234567:right}) does not generate any additional 
MI other that the scalar one, whose coefficient is $c_{1234567,0}$.

\paragraph{Coefficients.}
The $(2r+1)$ unknown coefficients $c_{1234567,i}$ 
can be determined by sampling Eq.(\ref{def:lad:Resi7:1234567:left}) and 
Eq.(\ref{def:lad:Resi7:1234567:right})
on $(2r+1)$ solutions of the 7fold-cut $(q_{(7)},k_{(7)})$
corresponding to $(2r+1)$ different values of $y_3$ 
(remember that $x_4$ also depends on $y_3$).
In this case, we finally find that only $c_{1234567,0}$ is non-vanishing,
\bea
c_{1234567,0} &=& - A^{\rm tree}(1^-,2^-,3^+,4^+) s_{12}^2 s_{23} \ , \\
c_{1234567,i} &=& 0 \qquad (1 \le i \le 2r+1) \ .
\eea
This means that, after reconstruction, 
$\Delta_{1234567}$ is constant,
$
\Delta_{1234567}(q,k) = c_{1234567,0} \ ,
$
which is exactly the result of Ref.\cite{Buchbinder:2005wp}.

\subsection{A contribution to the Crossed Amplitude}
\label{sec:4p:crossed}

\begin{figure}[h]
\vspace*{1.5cm}
\begin{center}
\hspace*{2.5cm}
\begin{minipage}{5cm}
%%%%%%%%%%%%%%%%%%%%%%%%%%%%%%%%%%%%%%%%
\begin{picture}(0,0)(0,0)
\SetScale{1.5}
\SetWidth{0.8}
\Line(0,15)(30,15)
\ArrowLine(0,15)(30,-15) \Text(8,+8)[]{{\fnsz{$k$}}}
\Line(30,-15)(0,-15)
\Line(30,15)(18,3) 
%\CArc(15,0)(7,225,45) 
\Line(12,-3)(0,-15)
\Line(0,15)(-30,15)
\ArrowLine(-30,15)(-30,-15) \Text(-60,0)[]{{\fnsz{$q$}}}
\Line(-30,-15)(0,-15)
\Line(30,15)(40,25)     \Text(65,45)[]{{\fnsz{$1^-$}}}
\Line(30,-15)(40,-25)   \Text(65,-45)[]{{\fnsz{$2^-$}}}
\Line(-30,15)(-40,25)     \Text(-65,45)[]{{\fnsz{$4^+$}}}
\Line(-30,-15)(-40,-25)   \Text(-65,-45)[]{{\fnsz{$3^+$}}}
\DashLine(15,20)(15,-20){2}
\DashLine(-15,20)(-15,-20){2}
%\DashLine(-35,0)(35,0){2}
\DashLine(-35,0)(-15,0){2} \DashLine(5,0)(25,0){2}
\Vertex(30,15){2}
\Vertex(30,-15){2}
\Vertex(-30,15){2}
\Vertex(-30,-15){2}
\Vertex(0,15){2}
\Vertex(0,-15){2}
\Text(35,27)[]{{\fnsz{$^+$}}}
\Text(10,27)[]{{\fnsz{$^-$}}}
\Text(-10,27)[]{{\fnsz{$^+$}}}
\Text(-35,27)[]{{\fnsz{$^-$}}}
\Text(35,-30)[]{{\fnsz{$^+$}}}
\Text(10,-30)[]{{\fnsz{$^-$}}}
\Text(-10,-30)[]{{\fnsz{$^+$}}}
\Text(-35,-30)[]{{\fnsz{$^-$}}}
\Text(50,15)[]{{\fnsz{$^-$}}}
\Text(50,-15)[]{{\fnsz{$^-$}}}
\Text(2,13)[]{{\fnsz{$^+$}}}
\Text(2,-15)[]{{\fnsz{$^+$}}}
\Text(-55,15)[]{{\fnsz{$^+$}}}
\Text(-55,-15)[]{{\fnsz{$^-$}}}
\end{picture}
%%%%%%%%%%%%%%%%%%%%%%%%%%%%%%%%%%%%%%%%
\end{minipage}
\begin{minipage}{3cm}
{\footnotesize
\bea
D_1 &=& k^2 \nn
D_2 &=& (k+p_2)^2 \nn
D_3 &=& (q+k-p_4)^2 \nn
D_4 &=& q^2 \nn
D_5 &=& (q+p_3)^2 \nn
D_6 &=& (q-p_4)^2 \nn
D_7 &=& (q+k+p_2+p_3)^2 \ . \nonumber
\eea
}
\end{minipage}
\end{center}
%\vspace*{1.0cm}
\caption{7fold-cut of the 4-point crossed diagram ($s$-channel)}
\label{fig:4p:crossed}
\end{figure}
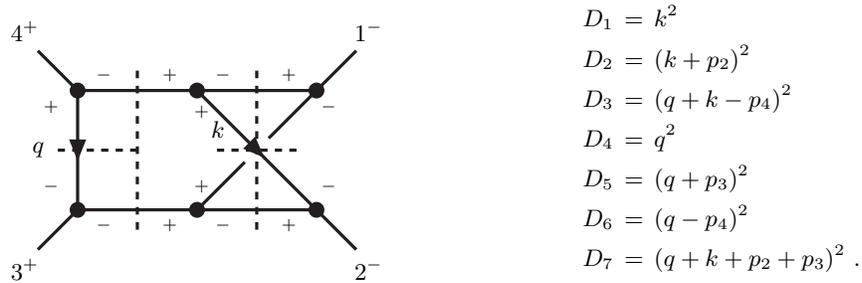

In this example, we consider the contribution to the (subleading-color) 
4-gluon MHV amplitude in ${\cal N}=4$ SYM coming from 
the helicity configuration depicted in Fig.\ref{fig:4p:crossed}, 
where only gluons circulate in both loops.
Its decomposition in terms of the crossed double-box MI 
in Fig.\ref{fig:4p:MIs} (right) is achieved by integrand-reduction.

\paragraph{On-Shell Solutions.}

The solutions of the 7fold-cut, 
$\db{1} = \ldots = \db{6} = \db{7} = 0$, can be decomposed according to 
Eqs.(\ref{eq:q_deco},\ref{eq:k_deco}), with the following definitions:
\bea
&& 
p_0^\mu = 0^\mu \ , \qquad
e_1^\mu = p_1^\mu \ , \qquad 
e_2^\mu = p_2^\mu \ , \qquad \\
\label{def:cros:ebasis7fold:1234567}
&& 
r_0^\mu = 0^\mu \ , \qquad
\tau_1^\mu = p_3^\mu \ , \qquad 
\tau_2^\mu = p_4^\mu\ . \qquad 
\label{def:cros:wbasis7fold:1234567}
\eea
As before, the on-shell conditions, 
$\db{1} = \ldots = \db{6} = \db{7} = 0$, cannot freeze the loop momenta,
and one component is left over as a free variable.
We choose $y_3$, namely the component of $k$ along $e_3$, as the variable
parametrizing the infinite set of solutions of this 7fold-cut.

\paragraph{Residue.}
The residue of this 7fold-cut is defined as,
\bea
\Delta_{1234567}(q,k) = 
\Res_{1234567}\{
N({q,k})
\}  \ ,
\label{def:cros:Resi7:1234567:left}
\eea
where $\Res_{1234567}\{N({q,k})\}$ is the product of the six 
3-point tree-amplitudes sitting in the vertices exposed by the cuts,
which can be built out of the tree-level expressions in Eq.(\ref{def:3p:tree})
used for the planar integrand.

To find the polynomial expression of $\Delta_{1234567}$, we use again 
our two criteria.
\begin{enumerate}
\item{\it Diagram topology and Vector basis.}

As for the planar case, in the considered 4-point double-box, 
only three external vectors are independent,
therefore any possible basis involving them
needs to be completed by an additional orthogonal vector $\omega_3$, 
previously defined in Eq.(\ref{eq:def:omega3}).

\item{\it Irreducible Scalar Products.} 

As in the planar case, the ISP's formed by the loop variables 
and the external momenta can be chosen to be
$(q\cdot p_2)$ and $(k \cdot p_3)$. 
Due to the explicit expression of the on-shell solutions, 
unlike the planar case (see Eq.(\ref{eq:4p:ladder:ISPs})), they both 
behave linearly in $y_3$.
This reflects the fact that these two ISP's are not linearly independent,
as one can explicitly see using the set of denominators in Fig.\ref{fig:4p:crossed}.

\end{enumerate}

\noindent
Therefore, $\Delta_{1234567}$ is a polynomial in $y_3$,
and can be parametrized as
\bea
\Delta_{1234567}(q,k) = 
\sum_{i=0}^r c_{1234567,i} \ ( \omega_3 \cdot k )^i 
=
\sum_{i=0}^r c_{1234567,i} \ \big((e_3\cdot p_3) \ y_3\big)^i \ ,
\label{def:cros:Resi7:1234567:right}
\eea
where $r$ is the maximum rank in the integration momenta.
The last equation holds because 
$(e_3 \cdot k) = (e_3 \cdot e_4) \ y_4$
and $y_4=0$, as required by the seven on-shell conditions.

\paragraph{Coefficients.} 
One can determine the $(r+1)$ unknown coefficients $c_{1234567,i}$,
by solving a system of equations obtained
from sampling Eq.(\ref{def:cros:Resi7:1234567:left}) and 
Eq.(\ref{def:cros:Resi7:1234567:right}) on $(r+1)$ solutions of the 7fold-cut.
The solution of the system finally reads,
\bea
c_{1234567,0} &=& - A^{\rm tree}(1^-,2^-,3^+,4^+) s_{12}^2 s_{23} \ , \\
c_{1234567,i} &\ne& 0 \qquad (i=1,\ldots,4) \ , \\
c_{1234567,j} &=& 0 \qquad (5 \le j \le r) \ .
\eea
By using these coefficients in the polynomial expression 
in Eq.(\ref{def:cros:Resi7:1234567:right}) we verify that 
the {\it local}-$(N=N)$ test is fulfilled.
In practice, we verify the equivalence of  
Eq.(\ref{def:cros:Resi7:1234567:left}) and 
the reconstructed polynomial Eq.(\ref{def:cros:Resi7:1234567:right})
when evaluated in any solution of the 7fold-cut other than the ones used
to determine the coefficients.

Notice that in this case the coefficients $c_{1234567,i}$ $(i=1,\ldots,4)$
are non-vanishing, but they do not affect the integrated result,
because they multiply spurious integrals that vanish upon integration.
Nevertheless their presence is important for 
the completeness of the polynomial expression in 
Eq.(\ref{def:cros:Resi7:1234567:right}).

\section{The MHV Pentabox in ${\cal N}=4$ SYM}

\begin{figure}[h]
%\vspace*{1.5cm}
\begin{center}
\hspace*{2.5cm}
\begin{minipage}{3.5cm}
%%%%%%%%%%%%%%%%%%%%%%%%%%%%%%%%%%%%%%%%
\begin{picture}(0,0)(0,0)
\SetScale{0.80}
\SetWidth{1.5}
\Line(-50,0)(-30, 25)
\ArrowLine(-50,0)(-30,-25) \Text(-40,-15)[]{{\fnsz{$q$}}}
\Line(-30, 25)(0, 15)
\Line(-30,-25)(0,-15)
\Line(0,15)(0,-15)
\Line(0,15)(30,15)
\ArrowLine(30,15)(30,-15) \Text(32,0)[]{{\fnsz{$k$}}}
\Line(30,-15)(0,-15)
\Line(30,15)(40,25)     \Text(38,25)[]{{\fnsz{$1$}}}
\Line(30,-15)(40,-25)   \Text(38,-25)[]{{\fnsz{$2$}}}
\Line(-30,-25)(-35,-35) \Text(-32,-35)[]{{\fnsz{$3$}}}
\Line(-60,0)(-50,0)     \Text(-53,0)[]{{\fnsz{$4$}}}
\Line(-30,+25)(-35,+35) \Text(-32,+35)[]{{\fnsz{$5$}}}
\end{picture}
%%%%%%%%%%%%%%%%%%%%%%%%%%%%%%%%%%%%%%%%
\end{minipage}
\begin{minipage}{3cm}
{\footnotesize
\bea
D_1 &=& k^2 \nn
D_2 &=& (k+p_2)^2 \nn
D_3 &=& (k-p_1)^2 \nn
D_4 &=& q^2 \nn
D_5 &=& (q+p_3)^2 \nn
D_6 &=& (q-p_4)^2 \nn
D_7 &=& (q-p_4-p_5)^2 \nn
D_8 &=& (q+k+p_2+p_3)^2 \ . \nonumber 
\eea
}
\end{minipage}
\end{center}
\caption{5-point pentabox diagram.}
\label{fig:5p:pentabox}
\end{figure}
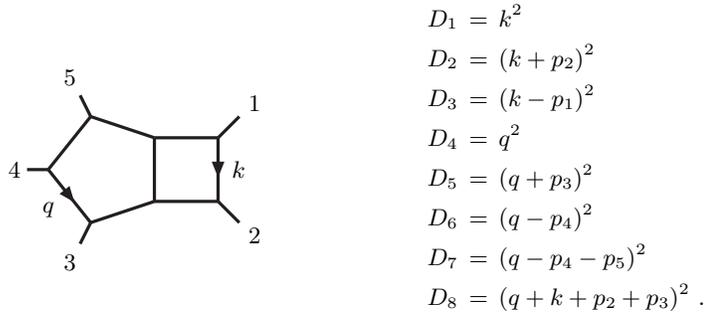

In the previous two examples we have discussed the 7fold-cut of a 4-point two-loop amplitude.
In this section we apply the integrand-reduction to the decomposition in terms of MI's of  
the 5-point two-loop pentabox in ${\cal N}=4$ SYM, shown in Fig.\ref{fig:5p:pentabox}.
The expression for the pentabox-integrand has been given in a very 
compact form in Ref.\cite{Carrasco:2011mn}.
The decomposition involves three types of contributions, coming from: 
a 5-point 8fold-cut, two 4-point 7fold-cuts, and two 5-point 7fold-cuts.

\subsection{Five-point Eightfold-Cut}
\label{sec:5p:12345678}

\begin{figure}[h]
\vspace*{1.0cm}
\begin{center}
%%%%%%%%%%%%%%%%%%%%%%%%%%%%%%%%%%%%%%%%
\begin{picture}(0,0)(0,0)
\SetScale{0.80}
\SetWidth{1.5}
\Line(-50,0)(-30, 25)
\ArrowLine(-50,0)(-30,-25) \Text(-40,-15)[]{{\fnsz{$q$}}}
\Line(-30, 25)(0, 15)
\Line(-30,-25)(0,-15)
\Line(0,15)(0,-15)
\Line(0,15)(30,15)
\ArrowLine(30,15)(30,-15) \Text(32,0)[]{{\fnsz{$k$}}}
\Line(30,-15)(0,-15)
\Line(30,15)(40,25)     \Text(38,25)[]{{\fnsz{$1$}}}
\Line(30,-15)(40,-25)   \Text(38,-25)[]{{\fnsz{$2$}}}
\Line(-30,-25)(-35,-35) \Text(-32,-35)[]{{\fnsz{$3$}}}
\Line(-60,0)(-50,0)     \Text(-53,0)[]{{\fnsz{$4$}}}
\Line(-30,+25)(-35,+35) \Text(-32,+35)[]{{\fnsz{$5$}}}
\DashLine(-22,0)(-42,+15){2}
\DashLine(-22,0)(-42,-15){2}
\DashLine(-22,0)(-13,+25){2}
\DashLine(-22,0)(-13,-25){2}
\DashLine(-22,0)(35,0){2}
\DashLine(15,0)(15,20){2}
\DashLine(15,0)(15,-20){2}
\end{picture}
%%%%%%%%%%%%%%%%%%%%%%%%%%%%%%%%%%%%%%%%
\end{center}
\vspace*{1.0cm}
\caption{5-point 8fold-cut $\Delta_{12345678}$.}
\label{fig:5p:8foldcut}
\end{figure}
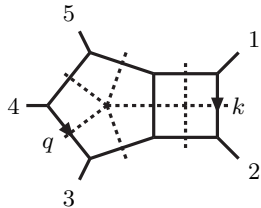

\paragraph{On-Shell Solutions.} 
The solutions of the 8fold-cut, 
$\db{1} = \ldots = \db{8} = 0$, depicted in Fig.\ref{fig:5p:8foldcut}, 
can be decomposed according to 
Eqs.(\ref{eq:q_deco},\ref{eq:k_deco}), with the following definitions:
\bea
&& 
r_0^\mu = 0^\mu \ , \qquad
e_1^\mu = p_1^\mu \ , \qquad 
e_2^\mu = p_2^\mu \ , \qquad \\
\label{def:ebasis8fold}
&& 
p_0^\mu = 0^\mu \ , \qquad
\tau_1^\mu = p_3^\mu \ , \qquad 
\tau_2^\mu = p_4^\mu \ . \qquad 
\label{def:wbasis8fold}
\eea
The eight on-shell conditions $\db{1} = \ldots = \db{8} = 0$ 
admit four solutions, where the loop momenta are completely frozen,
namely they are expressed in terms of external kinematic variables.

\paragraph{Residue.}
The residue of the 8fold-cut is defined as,
\bea
\Delta_{12345678}(q,k) = 
\Res_{12345678}\Big\{
{N({q,k})}
\Big\} 
\label{def:Resi8:left}
\eea
where the numerator function $N$ can be found in 
Table I (a) of \cite{Carrasco:2011mn},
and can be parametrized as 
\bea
\Delta_{12345678}(q,k) &=& 
c_{12345678,0} + 
c_{12345678,1} \ (q \cdot p_1) + \nn 
& & +
c_{12345678,2} \ (k \cdot p_4) + 
c_{12345678,3} \ (k \cdot p_5) \ .
\label{def:Resi8:right}
\eea
To derive its expression, we considered that 
this 5-point two-loop integral in four dimensions 
can depend on four external momenta,
hence four (out of five) legs can be chosen as a basis for 
decomposing the loop variables.
Moreover, this pentabox diagram admits three ISP's,
and we have chosen $(q \cdot p_1)$, $(k \cdot p_4)$, and $(k \cdot p_5)$.
Any other ISP can be expressed as a combination 
of them plus reducible scalar products.

The definition in Eq.(\ref{def:Resi8:right}) is compatible with the existence
of four MI's with denominators $\db{1}, \ldots, \db{8}$:
the scalar, plus three rank-1 integrals, each carrying one of the chosen ISP 
in the numerator.

\paragraph{Coefficients.} 
By sampling $N(q,k)$ at the four solutions of this 8fold-cut we can determine
the coefficients, and
find that $c_{12345678,0}$ and $c_{12345678,1}$ are non-vanishing,
while $c_{12345678,2} = c_{12345678,3} = 0.$
Therefore, only two (out of four) pentabox-like MI's will appear in the 
decomposition of the two-loop 5-point amplitude in ${\cal N}=4$ SYM.
Our result is in agreement with the results of 
Ref.~\cite{Bern:2006vw,Cachazo:2008vp}, although, in order to match
the coefficients given in these references, one must use an modified expression for 
$\Delta_{12345678}$, in which $c_{12345678,1}$ multiplies $((q-p_4-p_5) \cdot p_1)$.

\subsection{Five-point Sevenfold-Cut $(i)$}
\label{sec:cut7:1234568}

\begin{figure}[h]
\vspace*{1.0cm}
\begin{center}
%%%%%%%%%%%%%%%%%%%%%%%%%%%%%%%%%%%%%%%%
\begin{picture}(0,0)(0,0)
\SetScale{0.80}
\SetWidth{1.5}
\Line(0,15)(30,15)
\ArrowLine(30,15)(30,-15) \Text(32,0)[]{{\fnsz{$k$}}}
\Line(30,-15)(0,-15)
\Line(0,15)(0,-15)
\Line(0,15)(-30,15)
\ArrowLine(-30,15)(-30,-15) \Text(-32,0)[]{{\fnsz{$q$}}}
\Line(-30,-15)(0,-15)
\Line(30,15)(40,25)     \Text(38,25)[]{{\fnsz{$1$}}}
\Line(30,-15)(40,-25)   \Text(38,-25)[]{{\fnsz{$2$}}}
\Line(-30,15)(-40,25)     \Text(-38,25)[]{{\fnsz{$4$}}}
\Line(-30,-15)(-40,-25)   \Text(-38,-25)[]{{\fnsz{$3$}}}
\Line(0,+15)(0,30) \Text(0,32)[]{{\fnsz{$5$}}}
\DashLine(15,20)(15,-20){2}
\DashLine(-15,20)(-15,-20){2}
\DashLine(-35,0)(35,0){2}
\end{picture}
%%%%%%%%%%%%%%%%%%%%%%%%%%%%%%%%%%%%%%%%
\end{center}
\vspace*{1.0cm}
\caption{5-point 7fold-cut $\Delta_{1234568}$}
\label{fig:5p:7foldcut:1234568}
\end{figure}
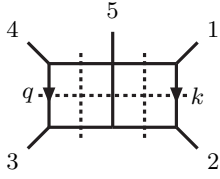

\paragraph{On-Shell Solutions.} 
The solutions of the 7fold-cut, 
$\db{1} = \ldots = \db{6} = \db{8} = 0$, in Fig.\ref{fig:5p:7foldcut:1234568}
can be decomposed according to 
Eqs.(\ref{eq:q_deco},\ref{eq:k_deco}), by using,
\bea
&& 
p_0^\mu = 0^\mu \ , \qquad
e_1^\mu = p_1^\mu \ , \qquad 
e_2^\mu = p_2^\mu \ , \qquad \\
\label{def:ebasis7fold:1234568}
&& 
r_0^\mu = 0^\mu \ , \qquad
\tau_1^\mu = p_3^\mu \ , \qquad 
\tau_2^\mu = p_4^\mu\ . \qquad 
\label{def:wbasis7fold:1234568}
\eea
In this case, the seven on-shell conditions, 
$\db{1} = \ldots = \db{6} = \db{8} = 0$, cannot freeze the loop momenta, and 
the generic solution reads,
\bea
q_{(7)}^\mu = x_4 \tau_4^\mu \ , \qquad k_{(7)}^\mu = y_3 e_3^\mu \ ,
\eea
where one of the on-shell cut-conditions 
imposes a non-linear relation among $x_4$ and $y_3$, which can be implicitly written as 
$x_4 = x_4(y_3)$.
We choose $y_3$, namely the component of $k$ along $e_3$, as the variable
parametrizing the infinite set of solutions of the 7fold-cut.

\paragraph{Residue.}

The residue of this 7fold-cut is defined as,
\bea
\Delta_{1234568}(q,k) = 
\Res_{1234568}\Bigg\{
{N({q,k}) - \Delta_{12345678}(q,k) \over \db{7}}
\Bigg\} \ .
\label{def:Resi7:1234568:left}
\eea
where $\Delta_{12345678}(q,k)$ is the polynomial residue of the 8fold-cut, reconstructed 
in Eq.(\ref{def:Resi8:right}).
To find the polynomial expression of $\Delta_{1234568}$, we use the following 
criteria.
\begin{enumerate}
\item{\it Diagram topology and Vector basis.}
We observe that 
this 5-point diagram depends on four external momenta 
and, as in the pentabox case,
its integrand does not contain any spurious ISP.

\item{\it Irreducible Scalar Products.} 
The ISP's which can be formed by the loop variables 
and the external momenta can be chosen to be
$(q\cdot p_1)$, $(q\cdot p_2)$, $(k \cdot p_3)$, $(k \cdot p_4)$. 
Due to the explicit expression of the on-shell solutions, 
on the cut, one has 
\bea
(q\cdot p_1) &\to& (q_{(7)} \cdot p_1) 
= - x_4 (\tau_4 \cdot p_1) \ , \\
(q\cdot p_2) &\to& (q_{(7)} \cdot p_2) 
= - x_4 (\tau_4 \cdot p_2) \ , \\
(k\cdot p_3) &\to& 
(k_{(7)} \cdot p_3) 
= y_3 (e_3 \cdot p_3) \ , \\
(k\cdot p_4) &\to& 
(k_{(7)} \cdot p_4) 
= y_3 (e_3 \cdot p_4) 
 \ .
\label{eq:5p:ladder:ISPs}
\eea
\end{enumerate}

\noindent
Therefore, $\Delta_{1234567}$ is polynomial in $x_4$ and $y_3$, and can be written as
\bea
\Delta_{1234568}(q,k) = 
c_{1234568,0} 
+ \sum_{i=1}^{r} c_{1234568,i}   \ (p_2 \cdot q)^i 
+ \sum_{i=1}^{r} c_{1234568,i+r} \ (p_3 \cdot k)^i \ ,
\label{def:Resi7:1234568:right}
\eea
where $r$ is the maximum rank in the integration momenta.

\paragraph{Coefficients.}
As by-now understood, the unknowns $c_{1234568,i}$ 
are found by sampling Eq.(\ref{def:Resi7:1234568:left}) 
and Eq.(\ref{def:Resi7:1234568:right}) 
on $(2r+1)$ solutions of the 7fold-cut. 
Also in this case we find that $\Delta_{1234568}(q,k)$
is a trivial polynomial, namely just a constant, because 
only $c_{1234568,0}$ is non-vanishing.

Let us finally remark that integrands with numerator $(p_2 \cdot q)^i$, and $(p_3 \cdot k)^i$ are non-vanishing, and would be MI's. They do not show up in this case because they are multiplied by null coefficients.

\subsection{Five-point Sevenfold-Cut $(ii)$}

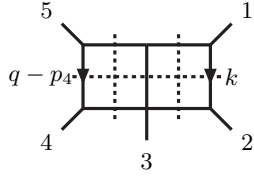
\begin{figure}[h]
\vspace*{1.0cm}
\begin{center}
%%%%%%%%%%%%%%%%%%%%%%%%%%%%%%%%%%%%%%%%
\begin{picture}(0,0)(0,0)
\SetScale{0.80}
\SetWidth{1.5}
\Line(0,15)(30,15)
\ArrowLine(30,15)(30,-15) \Text(32,0)[]{{\fnsz{$k$}}}
\Line(30,-15)(0,-15)
\Line(0,15)(0,-15)
\Line(0,15)(-30,15)
\ArrowLine(-30,15)(-30,-15) \Text(-40,0)[]{{\fnsz{$q-p_4$}}}
\Line(-30,-15)(0,-15) 
\Line(30,15)(40,25)     \Text(38,25)[]{{\fnsz{$1$}}}
\Line(30,-15)(40,-25)   \Text(38,-25)[]{{\fnsz{$2$}}}
\Line(0,-15)(0,-30) \Text(0,-32)[]{{\fnsz{$3$}}}
\Line(-30,-15)(-40,-25)   \Text(-38,-25)[]{{\fnsz{$4$}}}
\Line(-30,15)(-40,25)     \Text(-38,25)[]{{\fnsz{$5$}}}
\DashLine(15,20)(15,-20){2}
\DashLine(-15,20)(-15,-20){2}
\DashLine(-35,0)(35,0){2}
\end{picture}
%%%%%%%%%%%%%%%%%%%%%%%%%%%%%%%%%%%%%%%%
\end{center}
\vspace*{1.0cm}
\caption{5-point 7fold-cut $\Delta_{1234678}$}
\label{fig:5p:5fold:1234678}
\end{figure}
This case is specular to the previous one and can be treated by relabeling the 
external momenta $(1 \leftrightarrow 2, 3 \leftrightarrow 5)$.
As before only $c_{1234678,0}$ is non-vanishing, therefore only the scalar integral contribute.

\subsection{Four-point Sevenfold-Cut $(i)$}

\begin{figure}[h]
\vspace*{1.0cm}
\begin{center}
%%%%%%%%%%%%%%%%%%%%%%%%%%%%%%%%%%%%%%%%
\begin{picture}(0,0)(0,0)
\SetScale{0.80}
\SetWidth{1.5}
\Line(0,15)(30,15)
\ArrowLine(30,15)(30,-15) \Text(32,0)[]{{\fnsz{$k$}}}
\Line(30,-15)(0,-15)
\Line(0,15)(0,-15)
\Line(0,15)(-30,15)
\ArrowLine(-30,15)(-30,-15) \Text(-32,0)[]{{\fnsz{$q$}}}
\Line(-30,-15)(0,-15)
\Line(30,15)(40,25)     \Text(38,25)[]{{\fnsz{$1$}}}
\Line(30,-15)(40,-25)   \Text(38,-25)[]{{\fnsz{$2$}}}
\Line(-30,15)(-40,25)     \Text(-38,25)[]{{\fnsz{$4$}}}
\Line(-30,-15)(-40,-25)   \Text(-38,-25)[]{{\fnsz{$3$}}}
\Line(-30,+15)(-30,30) \Text(-25,32)[]{{\fnsz{$5$}}}
\DashLine(15,20)(15,-20){2}
\DashLine(-15,20)(-15,-20){2}
\DashLine(-35,0)(35,0){2}
\end{picture}
%%%%%%%%%%%%%%%%%%%%%%%%%%%%%%%%%%%%%%%%
\end{center}
\vspace*{1.0cm}
\caption{4-point 7fold-cut $\Delta_{1234578}$}
\label{fig:5p:7foldcut:1234578}
\end{figure}
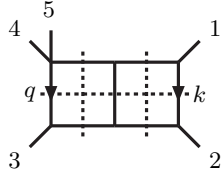

The 7fold-cut in Fig.~\ref{fig:5p:7foldcut:1234578} is equivalent to the ladder case 
treated in Sec. \ref{sec:4p:ladder}.
The solutions of 
$\db{1} = \ldots = \db{5} = \db{7} = \db{8} = 0$ can be decomposed according to 
Eqs.(\ref{eq:q_deco},\ref{eq:k_deco}), by using
\bea
&& 
r_0^\mu = 0^\mu \ , \qquad
e_1^\mu = p_1^\mu \ , \qquad 
e_2^\mu = p_2^\mu \ , \qquad \\
\label{def:ebasis7fold:1234578}
&& 
p_0^\mu = 0^\mu \ , \qquad
\tau_1^\mu = p_3^\mu \ , \qquad 
\tau_2^\mu = P_{45}^\mu - {s_{45}\over 2 P_{45} \cdot \tau_1} \tau_1^\mu \ . \qquad 
\label{def:wbasis7fold:1234578}
\eea
As in Sec. \ref{sec:4p:ladder}, the seven on-shell conditions
are not sufficient to freeze the loop momenta,
and one component is left over as a free variable.
We choose $y_3$, namely the component of $k$ along $e_3$, as the variable
parametrizing the infinite set of solutions.

The residue of the 7fold-cut, 
$\db{1} = \ldots = \db{5} = \db{7} = \db{8} = 0$, is defined as,
\bea
\Delta_{1234578}(q,k) = 
\Res_{1234578}\Bigg\{
{N({q,k}) - \Delta_{12345678}(q,k) \over \db{6}} 
\Bigg\} \ ,
\label{def:Resi7:1234578:left}
\eea
where $\Delta_{12345678}(q,k)$ is the polynomial residue of the 8fold-cut, reconstructed 
in Eq.(\ref{def:Resi8:right}).
The polynomial expression of $\Delta_{1234578}$ is equivalent to the one given 
in Eq.(\ref{def:lad:Resi7:1234567:right}),
where $\omega_3^\mu$, defined in Eq.(\ref{eq:def:omega3}), must be constructed with 
the basis vector $\{e_i\}$ 
used in Eq.(\ref{def:ebasis7fold:1234578}).

We determine the unknown coefficients and find
that only $c_{1234578,0}$ is non-vanishing.

\subsection{Four-point Sevenfold-Cut $(ii)$}

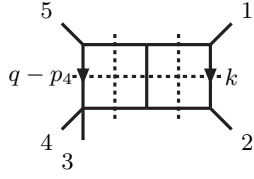
\begin{figure}[h]
\vspace*{1.0cm}
\begin{center}
%%%%%%%%%%%%%%%%%%%%%%%%%%%%%%%%%%%%%%%%
\begin{picture}(0,0)(0,0)
\SetScale{0.80}
\SetWidth{1.5}
\Line(0,15)(30,15)
\ArrowLine(30,15)(30,-15) \Text(32,0)[]{{\fnsz{$k$}}}
\Line(30,-15)(0,-15)
\Line(0,15)(0,-15)
\Line(0,15)(-30,15)
\ArrowLine(-30,15)(-30,-15) \Text(-40,0)[]{{\fnsz{$q-p_4$}}}
\Line(-30,-15)(0,-15) 
\Line(-30,-15)(-30,-30) \Text(-30,-32)[]{{\fnsz{$3$}}}
\Line(30,15)(40,25)     \Text(38,25)[]{{\fnsz{$1$}}}
\Line(30,-15)(40,-25)   \Text(38,-25)[]{{\fnsz{$2$}}}
\Line(-30,-15)(-40,-25)   \Text(-38,-25)[]{{\fnsz{$4$}}}
\Line(-30,15)(-40,25)     \Text(-38,25)[]{{\fnsz{$5$}}}
\DashLine(15,20)(15,-20){2}
\DashLine(-15,20)(-15,-20){2}
\DashLine(-35,0)(35,0){2}
\end{picture}
%%%%%%%%%%%%%%%%%%%%%%%%%%%%%%%%%%%%%%%%
\end{center}
\vspace*{1.0cm}
\caption{4-point 7fold-cut $\Delta_{1235678}$}
\label{fig:5p:7foldcut:1235678}
\end{figure}
Also the case in Fig.\ref{fig:5p:7foldcut:1235678} 
falls in the two-loop 4-point category discussed in Sec. \ref{sec:4p:ladder} 
and in the previous section.
The solutions of the 7fold-cut, 
$\db{1} = \ldots = \db{3} = \db{5} = \ldots = \db{8} = 0$, 
can be decomposed according to 
Eqs.(\ref{eq:q_deco},\ref{eq:k_deco}), with the following definitions:
\bea
&& 
p_0^\mu = 0^\mu \ , \qquad
e_1^\mu = p_1^\mu \ , \qquad 
e_2^\mu = p_2^\mu \ , \qquad \\
\label{def:ebasis7fold:1235678}
&& 
r_0^\mu = -p_4^\mu \ , \qquad
\tau_1^\mu = p_5^\mu \ , \qquad 
\tau_2^\mu = P_{34}^\mu - {s_{34}\over 2 P_{34} \cdot \tau_1} \tau_1^\mu\ . \qquad 
\label{def:wbasis7fold:1235678}
\eea
Again, we find it convenient to use $y_3$
as free variable to
parametrize the infinite set of solutions.

The residue, defined as
\bea
\Delta_{1235678}(q,k) = 
\Res_{1235678}\Bigg\{
{N({q,k}) - \Delta_{12345678}(q,k) \over \db{4}} 
\Bigg\} \ ,
\label{def:Resi7:1235678:left}
\eea
with $\Delta_{12345678}(q,k)$ being the the 8fold-cut polynomial
in Eq.(\ref{def:Resi8:right}),
can be written as in Eq.(\ref{def:lad:Resi7:1234567:right}),
where $\omega_3^\mu$ must be replaced by 
\bea
\omega_5^\mu = 
{ (p_5 \cdot e_3) e_4^\mu - (p_5 \cdot e_4) e_3^\mu \over e_3 \cdot e_4 } \ .
\label{eq:def:omega5}
\eea
By polynomial sampling we find that only $c_{1234578,0}$ is non-vanishing.

\subsection{Reconstructed Integrand}
Combining the results of the previous sections, 
the numerator of the planar 5-point pentabox diagram 
can be decomposed as,
\bea
N(q,k) &=&   \Delta_{12345678}(q,k)       +\nn &&
           + \Delta_{1234568}(q,k) \db{7}
           + \Delta_{1234578}(q,k) \db{6} +\nn &&
           + \Delta_{1234678}(q,k) \db{5}
           + \Delta_{1235678}(q,k) \db{4} =\nn
%&=& c_{12345678,0} +c_{12345678,1} \ ((q-p_4-p_5) \cdot p_1) +\nn &&
&=& c_{12345678,0} +c_{12345678,1} \ (q \cdot p_1) +\nn &&
           + c_{1234568,0} \db{7}
           + c_{1234578,0} \db{6} +\nn &&
           + c_{1234678,0} \db{5}
           + c_{1235678,0} \db{4} \ ,
\label{ed:Nreco:pentabox}
\eea
which corresponds to a decomposition of the integral in terms of six MI's,

\bea
%%%%%%%%%%%%%%%%%%%%%%%%%%%%%%%%%%%%%%%%
\begin{picture}(0,0)(0,0)
\SetScale{0.50}
\SetWidth{1.0}
\Line(-50,0)(-30, 25)
\ArrowLine(-50,0)(-30,-25) \Text(-25,-10)[]{{\tiny{$q$}}}
\Line(-30, 25)(0, 15)
\Line(-30,-25)(0,-15)
\Line(0,15)(0,-15)
\Line(0,15)(30,15)
\ArrowLine(30,15)(30,-15) \Text(20,0)[]{{\tiny{$k$}}}
\Line(30,-15)(0,-15)
\Line(30,15)(40,25)     \Text(22,18)[]{{\tiny{$1$}}}
\Line(30,-15)(40,-25)   \Text(22,-18)[]{{\tiny{$2$}}}
\Line(-30,-25)(-35,-35) \Text(-20,-22)[]{{\tiny{$3$}}}
\Line(-60,0)(-50,0)     \Text(-35,0)[]{{\tiny{$4$}}}
\Line(-30,+25)(-35,+35) \Text(-20,+22)[]{{\tiny{$5$}}}
 \Text(-11,0)[]{{\tiny{$N\!(q,\!k)$}}}
\end{picture}
%%%%%%%%%%%%%%%%%%%%%%%%%%%%%%%%%%%%%%%%
\hspace*{1cm}
& = &
c_{12345678,0}
\hspace*{1.5cm}
%%%%%%%%%%%%%%%%%%%%%%%%%%%%%%%%%%%%%%%%
\begin{picture}(0,0)(0,0)
\SetScale{0.50}
\SetWidth{1.0}
\Line(-50,0)(-30, 25)
\ArrowLine(-50,0)(-30,-25) %\Text(-25,-10)[]{{\tiny{$q$}}}
\Line(-30, 25)(0, 15)
\Line(-30,-25)(0,-15)
\Line(0,15)(0,-15)
\Line(0,15)(30,15)
\ArrowLine(30,15)(30,-15) %\Text(20,0)[]{{\tiny{$k$}}}
\Line(30,-15)(0,-15)
\Line(30,15)(40,25)     \Text(22,18)[]{{\tiny{$1$}}}
\Line(30,-15)(40,-25)   \Text(22,-18)[]{{\tiny{$2$}}}
\Line(-30,-25)(-35,-35) \Text(-20,-22)[]{{\tiny{$3$}}}
\Line(-60,0)(-50,0)     \Text(-35,0)[]{{\tiny{$4$}}}
\Line(-30,+25)(-35,+35) \Text(-20,+22)[]{{\tiny{$5$}}}
\end{picture}
%%%%%%%%%%%%%%%%%%%%%%%%%%%%%%%%%%%%%%%%
\hspace*{1.0cm}
+ c_{12345678,1}
\hspace*{1.5cm}
%%%%%%%%%%%%%%%%%%%%%%%%%%%%%%%%%%%%%%%%
\begin{picture}(0,0)(0,0)
\SetScale{0.50}
\SetWidth{1.0}
\Line(-50,0)(-30, 25)
\ArrowLine(-50,0)(-30,-25) %\Text(-40,-15)[]{{\fnsz{$q$}}}
\Line(-30, 25)(0, 15)
\Line(-30,-25)(0,-15)
\Line(0,15)(0,-15)
\Line(0,15)(30,15)
\ArrowLine(30,15)(30,-15) %\Text(32,0)[]{{\fnsz{$k$}}}
\Line(30,-15)(0,-15)
\Line(30,15)(40,25)     \Text(22,18)[]{{\tiny{$1$}}}
\Line(30,-15)(40,-25)   \Text(22,-18)[]{{\tiny{$2$}}}
\Line(-30,-25)(-35,-35) \Text(-20,-22)[]{{\tiny{$3$}}}
\Line(-60,0)(-50,0)     \Text(-35,0)[]{{\tiny{$4$}}}
\Line(-30,+25)(-35,+35) \Text(-20,+22)[]{{\tiny{$5$}}}
 \Text(-11,0)[]{{\tiny{$(q \! \cdot \!p_1\!)$}}}
\end{picture}
%%%%%%%%%%%%%%%%%%%%%%%%%%%%%%%%%%%%%%%%
\hspace*{1cm} + 
\nn && \nn && \nn && 
+c_{1234568,0}
\hspace*{1.0cm}
%%%%%%%%%%%%%%%%%%%%%%%%%%%%%%%%%%%%%%%%
\begin{picture}(0,0)(0,0)
\SetScale{0.50}
\SetWidth{1.0}
\Line(0,15)(30,15)
\ArrowLine(30,15)(30,-15) %\Text(32,0)[]{{\fnsz{$k$}}}
\Line(30,-15)(0,-15)
\Line(0,15)(0,-15)
\Line(0,15)(-30,15)
\ArrowLine(-30,15)(-30,-15) %\Text(-32,0)[]{{\fnsz{$q$}}}
\Line(-30,-15)(0,-15)
\Line(30,15)(40,25)     \Text(22,18)[]{{\tiny{$1$}}}
\Line(30,-15)(40,-25)   \Text(22,-18)[]{{\tiny{$2$}}}
\Line(-30,15)(-40,25)   \Text(-22,18)[]{{\tiny{$4$}}}
\Line(-30,-15)(-40,-25) \Text(-22,-18)[]{{\tiny{$3$}}}
\Line(0,+15)(0,30) \Text(0,22)[]{{\tiny{$5$}}}
\end{picture}
%%%%%%%%%%%%%%%%%%%%%%%%%%%%%%%%%%%%%%%%
\hspace*{1.0cm}
+c_{1234578,0}
\hspace*{1.0cm}
%%%%%%%%%%%%%%%%%%%%%%%%%%%%%%%%%%%%%%%%
\begin{picture}(0,0)(0,0)
\SetScale{0.50}
\SetWidth{1.0}
\Line(0,15)(30,15)
\ArrowLine(30,15)(30,-15) %\Text(32,0)[]{{\fnsz{$k$}}}
\Line(30,-15)(0,-15)
\Line(0,15)(0,-15)
\Line(0,15)(-30,15)
\ArrowLine(-30,15)(-30,-15) %\Text(-32,0)[]{{\fnsz{$q$}}}
\Line(-30,-15)(0,-15)
\Line(30,15)(40,25)     \Text(22,18)[]{{\tiny{$1$}}}
\Line(30,-15)(40,-25)   \Text(22,-18)[]{{\tiny{$2$}}}
\Line(-30,15)(-40,25)   \Text(-22,18)[]{{\tiny{$4$}}}
\Line(-30,-15)(-40,-25) \Text(-22,-18)[]{{\tiny{$3$}}}
\Line(-30,+15)(-30,30)  \Text(-15,20)[]{{\tiny{$5$}}}
\end{picture}
%%%%%%%%%%%%%%%%%%%%%%%%%%%%%%%%%%%%%%%%
\hspace*{1cm} + 
\nn && \nn && \nn && 
+c_{1234678,0}
\hspace*{1cm} 
%%%%%%%%%%%%%%%%%%%%%%%%%%%%%%%%%%%%%%%%
\begin{picture}(0,0)(0,0)
\SetScale{0.50}
\SetWidth{1.0}
\Line(0,15)(30,15)
\ArrowLine(30,15)(30,-15) %\Text(32,0)[]{{\fnsz{$k$}}}
\Line(30,-15)(0,-15)
\Line(0,15)(0,-15)
\Line(0,15)(-30,15)
\ArrowLine(-30,15)(-30,-15) %\Text(-40,0)[]{{\fnsz{$q-p_4$}}}
\Line(-30,-15)(0,-15) 
\Line(30,15)(40,25)     \Text(22,18)[]{{\tiny{$1$}}}
\Line(30,-15)(40,-25)   \Text(22,-18)[]{{\tiny{$2$}}}
\Line(0,-15)(0,-30)     \Text(0,-20)[]{{\tiny{$3$}}}
\Line(-30,-15)(-40,-25) \Text(-22,-18)[]{{\tiny{$4$}}}
\Line(-30,15)(-40,25)   \Text(-22,18)[]{{\tiny{$5$}}}
\end{picture}
%%%%%%%%%%%%%%%%%%%%%%%%%%%%%%%%%%%%%%%%
\hspace*{1.0cm}
+c_{1235678,0}
\hspace*{1.0cm}
%%%%%%%%%%%%%%%%%%%%%%%%%%%%%%%%%%%%%%%%
\begin{picture}(0,0)(0,0)
\SetScale{0.50}
\SetWidth{1.0}
\Line(0,15)(30,15)
\ArrowLine(30,15)(30,-15) %\Text(32,0)[]{{\fnsz{$k$}}}
\Line(30,-15)(0,-15)
\Line(0,15)(0,-15)
\Line(0,15)(-30,15)
\Line(-30,15)(-30,-15) %\Text(-40,0)[]{{\fnsz{$q-p_4$}}}
\Line(-30,-15)(0,-15) 
\Line(-30,-15)(-30,-30) \Text(-15,-20)[]{{\tiny{$3$}}}
\Line(30,15)(40,25)     \Text(22,18)[]{{\tiny{$1$}}}
\Line(30,-15)(40,-25)   \Text(22,-18)[]{{\tiny{$2$}}}
\Line(-30,-15)(-40,-25) \Text(-22,-18)[]{{\tiny{$4$}}}
\Line(-30,15)(-40,25)   \Text(-22,18)[]{{\tiny{$5$}}}
\end{picture}
%%%%%%%%%%%%%%%%%%%%%%%%%%%%%%%%%%%%%%%%
\\ && \nn && \nonumber
\label{eq:pentabox:reduction}
\eea
\noindent
We checked the correctness of the Eq.(\ref{ed:Nreco:pentabox})
through the {\it global}-$(N=N)$ test, namely by verifying the identity
of the {\it l.h.s} and of the {\it r.h.s.} for arbitrary values of 
$q$ and $k$.
%In particular one can notice from the {\it r.h.s.} of Eq.(\ref{ed:Nreco:pentabox}) 
%that $N(q,k)$ does not depend on $k$, {\it i.e.} $N(q,k)=N(q)$, 
%because $D_4$, $D_5$, $D_6$, $D_7$ contain only $q$. 

\section{The MHV Pentacross in ${\cal N}=4$ SYM}

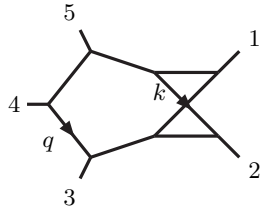
\begin{figure}[h]
\vspace*{1.5cm}
\begin{center}
\hspace*{2.5cm}
\begin{minipage}{3.5cm}
%%%%%%%%%%%%%%%%%%%%%%%%%%%%%%%%%%%%%%%%
\begin{picture}(0,0)(0,0)
\SetScale{0.80}
\SetWidth{1.5}
\Line(-50,0)(-30, 25)
\ArrowLine(-50,0)(-30,-25) \Text(-40,-15)[]{{\fnsz{$q$}}}
\Line(-30, 25)(0, 15)
\Line(-30,-25)(0,-15)
\Line(30,15)(0,-15)
\Line(0,15)(30,15)
\ArrowLine(0,15)(30,-15) \Text(2,+5)[]{{\fnsz{$k$}}}
\Line(30,-15)(0,-15)
\Line(30,15)(40,25)     \Text(38,25)[]{{\fnsz{$1$}}}
\Line(30,-15)(40,-25)   \Text(38,-25)[]{{\fnsz{$2$}}}
\Line(-30,-25)(-35,-35) \Text(-32,-35)[]{{\fnsz{$3$}}}
\Line(-60,0)(-50,0)     \Text(-53,0)[]{{\fnsz{$4$}}}
\Line(-30,+25)(-35,+35) \Text(-32,+35)[]{{\fnsz{$5$}}}
\end{picture}
%%%%%%%%%%%%%%%%%%%%%%%%%%%%%%%%%%%%%%%%
\end{minipage}
\begin{minipage}{3cm}
{\footnotesize
\bea
D_1 &=& k^2 \nn
D_2 &=& (k+p_2)^2 \nn
D_3 &=& (k+q-p_4-p_5)^2 \nn
D_4 &=& q^2 \nn
D_5 &=& (q+p_3)^2 \nn
D_6 &=& (q-p_4)^2 \nn
D_7 &=& (q-p_4-p_5)^2 \nn
D_8 &=& (q+k+p_2+p_3)^2 \ . \nonumber
\eea
}
\end{minipage}
\end{center}
\caption{5-point Pentacross (non-planar)}
\label{fig:5p:pentacross}
\end{figure}

In this section we present the result of the integrand
reduction of the MHV pentacross diagram in ${\cal N}=4$ SYM,
depicted in Fig.\ref{fig:5p:pentacross}.
The expression for the its integrand has been given in Table I (b) of 
Ref.\cite{Carrasco:2011mn}, and happens to have the same expression of 
the planar diagram, although it is sitting on a different set of denominators, listed in 
Fig.\ref{fig:5p:pentacross}.

The integrand-reduction follows the same pattern as described in the Section 3 and 4.
In particular the expressions of the polynomial residues of the 
5-point $8-$fold cut,
and 4-point $7-$fold cut, can be obtained following the same procedures as 
in Sec.~\ref{sec:5p:12345678} and Sec.~\ref{sec:4p:crossed}, respectively.
The 5-point $7-$fold cut deserves a dedicated discussion.

\subsection{Five-point Sevenfold-Cut }
\label{sec:cut7:1234568}

\begin{figure}[h]
\vspace*{1.5cm}
\begin{center}
%%%%%%%%%%%%%%%%%%%%%%%%%%%%%%%%%%%%%%%%
\begin{picture}(0,0)(0,0)
\SetScale{0.80}
\SetWidth{1.5}
\Line(0,15)(30,15)
\ArrowLine(0,15)(30,-15) \Text(32,0)[]{{\fnsz{$k$}}}
\Line(30,-15)(0,-15)
\Line(30,15)(0,-15)
\Line(0,15)(-30,15)
\ArrowLine(-30,15)(-30,-15) \Text(-32,0)[]{{\fnsz{$q$}}}
\Line(-30,-15)(0,-15)
\Line(30,15)(40,25)     \Text(38,25)[]{{\fnsz{$1$}}}
\Line(30,-15)(40,-25)   \Text(38,-25)[]{{\fnsz{$2$}}}
\Line(-30,15)(-40,25)     \Text(-38,25)[]{{\fnsz{$4$}}}
\Line(-30,-15)(-40,-25)   \Text(-38,-25)[]{{\fnsz{$3$}}}
\Line(0,+15)(0,30) \Text(0,32)[]{{\fnsz{$5$}}}
\DashLine(15,20)(15,-20){2}
\DashLine(-15,20)(-15,-20){2}
\DashLine(-35,0)(-15,0){2}
\DashLine(5,0)(25,0){2}
\end{picture}
%%%%%%%%%%%%%%%%%%%%%%%%%%%%%%%%%%%%%%%%
\end{center}
\vspace*{0.5cm}
\caption{5-point 7fold-cut $\Delta_{1234568}$}
\label{fig:5p:7foldcut:cross:1234568}
\end{figure}
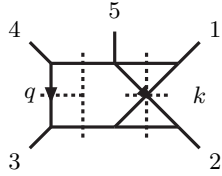

\paragraph{On-Shell Solutions.} 
The solutions of the 5point 7fold-cut, 
$\db{1} = \ldots = \db{6} = \db{8} = 0$, depicted in Fig.~\ref{fig:5p:7foldcut:cross:1234568}
can be decomposed according to 
Eqs.(\ref{eq:q_deco},\ref{eq:k_deco}), with the following definitions:
\bea
&& 
p_0^\mu = 0^\mu \ , \qquad
e_1^\mu = p_1^\mu \ , \qquad 
e_2^\mu = p_2^\mu \ , \qquad \\
\label{def:cross:ebasis7fold:1234568}
&& 
r_0^\mu = 0^\mu \ , \qquad
\tau_1^\mu = p_3^\mu \ , \qquad 
\tau_2^\mu = p_4^\mu\ . \qquad 
\label{def:cross:wbasis7fold:1234568}
\eea
The on-shell conditions, 
$\db{1} = \ldots = \db{6} = \db{8} = 0$, cannot freeze the loop momenta, and 
we choose $y_3$ to parametrize the infinite set of solutions.

\paragraph{Residue.}

The residue is defined as,
\bea
\Delta_{1234568}(q,k) = 
\Res_{1234568}\Bigg\{
{N({q,k}) - \Delta_{12345678}(q,k) \over \db{7}}
\Bigg\} \ .
\label{def:cross:Resi7:1234568:left}
\eea
The function $\Delta_{12345678}$ is the 8fold-cut polynomial of the pentacross.
To find the polynomial expression of $\Delta_{1234568}$, we use the following 
criteria.
\begin{enumerate}
\item{\it Diagram topology and Vector basis.}
We observe that 
this 5-point diagram depends on four external momenta,
therefore
its integrand does not contain any spurious ISP.

\item{\it Irreducible Scalar Products.} 
The ISP's formed by the loop variables 
and the external momenta can be chosen to be
$(q\cdot p_1)$, $(q\cdot p_2)$, $(k \cdot p_3)$, $(k \cdot p_4)$. 
Due to the explicit expression of the on-shell solutions, 
they all behave linearly in $y_3$. 
This reflects the fact that although irreducible, these four ISP's 
are not linearly independent.
\end{enumerate}

\noindent
Therefore, $\Delta_{1234568}$ is as a polynomial in $y_3$, and can be written as
\bea
\Delta_{1234568}(q,k) = 
c_{1234568,0} 
+ \sum_{i=1}^{r} c_{1234568,i}   \ (p_1 \cdot q)^i \ ,
\label{def:cross:Resi7:1234568:right}
\eea
where $r$ is the maximum rank in the integration momenta.

\paragraph{Coefficients.}
As by-now understood, the unknowns $c_{1234568,i}$ 
are found by sampling Eq.(\ref{def:cross:Resi7:1234568:left}) 
and Eq.(\ref{def:cross:Resi7:1234568:right}) 
on $(r+1)$ solutions of the 7fold-cut.
Also in this case we find that 
$c_{1234568,0}$ is the only non-vanishing coefficient.

Let us finally remark that integrands with numerator $(p_1 \cdot q)^i$ 
would generate MI's. They do not show up in the final result because 
they are multiplied by null coefficients.

\subsection{Reconstructed Integrand}

The integrand-decomposition for the MHV pentacross diagram 
given in Table I (b) of Ref. \cite{Carrasco:2011mn}
has the same structure of the planar pentabox,
\bea
N(q,k) &=&   \Delta_{12345678}(q,k)       +\nn &&
           + \Delta_{1234568}(q,k) \db{7}
           + \Delta_{1234578}(q,k) \db{6} +\nn &&
           + \Delta_{1234678}(q,k) \db{5}
           + \Delta_{1235678}(q,k) \db{4} =\nn
&=& c_{12345678,0} +c_{12345678,1} \ (q \cdot p_1) +\nn &&
           + c_{1234568,0} \db{7}
           + c_{1234578,0} \db{6} +\nn &&
           + c_{1234678,0} \db{5}
           + c_{1235678,0} \db{4} \ ,
\label{ed:Nreco:pentacross}
\eea
and contains the same coefficients of Eq.(\ref{ed:Nreco:pentabox}).
The above decomposition has been verified to fulfill the {\it global}-$(N=N)$ test.
Therefore, the final expression of the pentacross diagram
in terms of MI's reads,

\bea
%%%%%%%%%%%%%%%%%%%%%%%%%%%%%%%%%%%%%%%%
\begin{picture}(0,0)(0,0)
\SetScale{0.50}
\SetWidth{1.0}
\Line(-50,0)(-30, 25)
\ArrowLine(-50,0)(-30,-25) \Text(-25,-10)[]{{\tiny{$q$}}}
\Line(-30, 25)(0, 15)
\Line(-30,-25)(0,-15)
\Line(30,15)(0,-15)
\Line(0,15)(30,15)
\ArrowLine(0,15)(30,-15) \Text(15,-3)[]{{\tiny{$k$}}}
\Line(30,-15)(0,-15)
\Line(30,15)(40,25)     \Text(22,18)[]{{\tiny{$1$}}}
\Line(30,-15)(40,-25)   \Text(22,-18)[]{{\tiny{$2$}}}
\Line(-30,-25)(-35,-35) \Text(-20,-22)[]{{\tiny{$3$}}}
\Line(-60,0)(-50,0)     \Text(-35,0)[]{{\tiny{$4$}}}
\Line(-30,+25)(-35,+35) \Text(-20,+22)[]{{\tiny{$5$}}}
 \Text(-11,0)[]{{\tiny{$N\!(q,\!k)$}}}
\end{picture}
%%%%%%%%%%%%%%%%%%%%%%%%%%%%%%%%%%%%%%%%
\hspace*{1cm}
& = &
c_{12345678,0}
\hspace*{1.5cm}
%%%%%%%%%%%%%%%%%%%%%%%%%%%%%%%%%%%%%%%%
\begin{picture}(0,0)(0,0)
\SetScale{0.50}
\SetWidth{1.0}
\Line(-50,0)(-30, 25)
\ArrowLine(-50,0)(-30,-25) %\Text(-25,-10)[]{{\tiny{$q$}}}
\Line(-30, 25)(0, 15)
\Line(-30,-25)(0,-15)
\Line(30,15)(0,-15)
\Line(0,15)(30,15)
\ArrowLine(0,15)(30,-15) %\Text(20,0)[]{{\tiny{$k$}}}
\Line(30,-15)(0,-15)
\Line(30,15)(40,25)     \Text(22,18)[]{{\tiny{$1$}}}
\Line(30,-15)(40,-25)   \Text(22,-18)[]{{\tiny{$2$}}}
\Line(-30,-25)(-35,-35) \Text(-20,-22)[]{{\tiny{$3$}}}
\Line(-60,0)(-50,0)     \Text(-35,0)[]{{\tiny{$4$}}}
\Line(-30,+25)(-35,+35) \Text(-20,+22)[]{{\tiny{$5$}}}
\end{picture}
%%%%%%%%%%%%%%%%%%%%%%%%%%%%%%%%%%%%%%%%
\hspace*{1.0cm}
+ c_{12345678,1}
\hspace*{1.5cm}
%%%%%%%%%%%%%%%%%%%%%%%%%%%%%%%%%%%%%%%%
\begin{picture}(0,0)(0,0)
\SetScale{0.50}
\SetWidth{1.0}
\Line(-50,0)(-30, 25)
\ArrowLine(-50,0)(-30,-25) %\Text(-40,-15)[]{{\fnsz{$q$}}}
\Line(-30, 25)(0, 15)
\Line(-30,-25)(0,-15)
\Line(30,15)(0,-15)
\Line(0,15)(30,15)
\ArrowLine(0,15)(30,-15) %\Text(32,0)[]{{\fnsz{$k$}}}
\Line(30,-15)(0,-15)
\Line(30,15)(40,25)     \Text(22,18)[]{{\tiny{$1$}}}
\Line(30,-15)(40,-25)   \Text(22,-18)[]{{\tiny{$2$}}}
\Line(-30,-25)(-35,-35) \Text(-20,-22)[]{{\tiny{$3$}}}
\Line(-60,0)(-50,0)     \Text(-35,0)[]{{\tiny{$4$}}}
\Line(-30,+25)(-35,+35) \Text(-20,+22)[]{{\tiny{$5$}}}
 \Text(-11,0)[]{{\tiny{$(q \! \cdot \!p_1\!)$}}}
\end{picture}
%%%%%%%%%%%%%%%%%%%%%%%%%%%%%%%%%%%%%%%%
\hspace*{1cm} + 
\nn && \nn && \nn && 
+c_{1234568,0}
\hspace*{1.0cm}
%%%%%%%%%%%%%%%%%%%%%%%%%%%%%%%%%%%%%%%%
\begin{picture}(0,0)(0,0)
\SetScale{0.50}
\SetWidth{1.0}
\Line(0,15)(30,15)
\ArrowLine(0,15)(30,-15) %\Text(32,0)[]{{\fnsz{$k$}}}
\Line(30,-15)(0,-15)
\Line(30,15)(0,-15)
\Line(0,15)(-30,15)
\ArrowLine(-30,15)(-30,-15) %\Text(-32,0)[]{{\fnsz{$q$}}}
\Line(-30,-15)(0,-15)
\Line(30,15)(40,25)     \Text(22,18)[]{{\tiny{$1$}}}
\Line(30,-15)(40,-25)   \Text(22,-18)[]{{\tiny{$2$}}}
\Line(-30,15)(-40,25)   \Text(-22,18)[]{{\tiny{$4$}}}
\Line(-30,-15)(-40,-25) \Text(-22,-18)[]{{\tiny{$3$}}}
\Line(0,+15)(0,30) \Text(0,22)[]{{\tiny{$5$}}}
\end{picture}
%%%%%%%%%%%%%%%%%%%%%%%%%%%%%%%%%%%%%%%%
\hspace*{1.0cm}
+c_{1234578,0}
\hspace*{1.0cm}
%%%%%%%%%%%%%%%%%%%%%%%%%%%%%%%%%%%%%%%%
\begin{picture}(0,0)(0,0)
\SetScale{0.50}
\SetWidth{1.0}
\Line(0,15)(30,15)
\ArrowLine(0,15)(30,-15) %\Text(32,0)[]{{\fnsz{$k$}}}
\Line(30,-15)(0,-15)
\Line(30,15)(0,-15)
\Line(0,15)(-30,15)
\ArrowLine(-30,15)(-30,-15) %\Text(-32,0)[]{{\fnsz{$q$}}}
\Line(-30,-15)(0,-15)
\Line(30,15)(40,25)     \Text(22,18)[]{{\tiny{$1$}}}
\Line(30,-15)(40,-25)   \Text(22,-18)[]{{\tiny{$2$}}}
\Line(-30,15)(-40,25)   \Text(-22,18)[]{{\tiny{$4$}}}
\Line(-30,-15)(-40,-25) \Text(-22,-18)[]{{\tiny{$3$}}}
\Line(-30,+15)(-30,30)  \Text(-15,20)[]{{\tiny{$5$}}}
\end{picture}
%%%%%%%%%%%%%%%%%%%%%%%%%%%%%%%%%%%%%%%%
\hspace*{1.0cm}
\\ && \nn \nn && \nonumber
+c_{1234678,0}
\hspace*{1cm} 
%%%%%%%%%%%%%%%%%%%%%%%%%%%%%%%%%%%%%%%%
\begin{picture}(0,0)(0,0)
\SetScale{0.50}
\SetWidth{1.0}
\Line(0,15)(30,15)
\ArrowLine(0,15)(30,-15) %\Text(32,0)[]{{\fnsz{$k$}}}
\Line(30,-15)(0,-15)
\Line(30,15)(0,-15)
\Line(0,15)(-30,15)
\ArrowLine(-30,15)(-30,-15) %\Text(-40,0)[]{{\fnsz{$q-p_4$}}}
\Line(-30,-15)(0,-15) 
\Line(30,15)(40,25)     \Text(22,18)[]{{\tiny{$1$}}}
\Line(30,-15)(40,-25)   \Text(22,-18)[]{{\tiny{$2$}}}
\Line(0,-15)(0,-30)     \Text(0,-20)[]{{\tiny{$3$}}}
\Line(-30,-15)(-40,-25) \Text(-22,-18)[]{{\tiny{$4$}}}
\Line(-30,15)(-40,25)   \Text(-22,18)[]{{\tiny{$5$}}}
\end{picture}
%%%%%%%%%%%%%%%%%%%%%%%%%%%%%%%%%%%%%%%%
\hspace*{1cm} 
+c_{1235678,0}
\hspace*{1.0cm}
%%%%%%%%%%%%%%%%%%%%%%%%%%%%%%%%%%%%%%%%
\begin{picture}(0,0)(0,0)
\SetScale{0.50}
\SetWidth{1.0}
\Line(0,15)(30,15)
\ArrowLine(0,15)(30,-15) %\Text(32,0)[]{{\fnsz{$k$}}}
\Line(30,-15)(0,-15)
\Line(30,15)(0,-15)
\Line(0,15)(-30,15)
\Line(-30,15)(-30,-15) %\Text(-40,0)[]{{\fnsz{$q-p_4$}}}
\Line(-30,-15)(0,-15) 
\Line(-30,-15)(-30,-30) \Text(-15,-20)[]{{\tiny{$3$}}}
\Line(30,15)(40,25)     \Text(22,18)[]{{\tiny{$1$}}}
\Line(30,-15)(40,-25)   \Text(22,-18)[]{{\tiny{$2$}}}
\Line(-30,-15)(-40,-25) \Text(-22,-18)[]{{\tiny{$4$}}}
\Line(-30,15)(-40,25)   \Text(-22,18)[]{{\tiny{$5$}}}
\end{picture}
%%%%%%%%%%%%%%%%%%%%%%%%%%%%%%%%%%%%%%%%
\\
&& \nonumber 
\label{eq:pentacross:reduction}
\eea

\section{Conclusions}
We illustrated a first implementation of the integrand-reduction method 
for two-loop scattering amplitudes.
We have shown that the residues of the amplitudes on the multi-particle cuts 
are polynomials written in terms of independent irreducible scalar products 
formed by loop momenta and either external momenta or polarization vectors 
built out of them.
The independence conditions among irreducible scalar products can be 
investigated through their polynomial behavior in terms of the components 
of the loop momenta still undetermined after imposing the on-shell cut-conditions.

The reduction of the amplitudes in terms of master integrals can 
be realized through polynomial fitting of the integrand, 
without any need of an apriori knowledge of the integral basis.
We discussed how the polynomial shapes of the residues determine the basis
of master integrals appearing in the final result.
In particular, we have found that 
the multiparticle residues of amplitudes with less then five 
external legs can eventually be written in terms of 
spurious irreducible scalar products which do not generate any master integral 
upon integration.

We applied the integrand-reduction algorithm 
to cases of modest complexity, such as 
planar and non-planar contributions 
to the 4-point MHV and 5-point MHV amplitudes in ${\cal N}=4$ SYM.
We worked out the polynomials parametrizing the residues at the 
4-point 7fold-cut,
5-point 7fold-cut,
5-point 8fold-cut which contribute to the decomposition of the considered amplitudes.

In this work we identified general principles 
which could guide
in the classification of all polynomial residues 
required by the complete integrand-reduction of arbitrary two-loop amplitudes. 
The technique we have presented extends the well-established analogous 
method for one-loop amplitudes,
and can be considered a preliminary study towards the systematic 
reduction  at the integrand-level of multi-loop amplitudes in any gauge theory,
suitable for their automated semianalytic evaluation.
%% basme'l ql

\section*{Acknowledgments}

We would like to thank John Joseph Carrasco for providing expressions 
for the integrands used in Sections~4 and 5,
and for feedback on the manuscript.
We also thank Christian Sturm and Francesco Tramontano for interesting 
discussions, and Gudrun Heinrich for comments on the manuscript. \\
G.O. wishes to acknowledge the kind hospitality of the 
Max-Planck Insitut f\"ur Physik in Munich
at several stages during the completion of this project. \\
The work of P.M. was supported by the Sofja Kovaleskaja Award Grant of 
the Alexander von Humboldt Stiftung, founded by the German Federal Ministry 
of Education and Research, and by the Ramon y Cajal Fellowship Program. 
The work of G.O. was supported in part by the National Science Foundation 
Grant PHY-0855489.

\bibliographystyle{utphys} 
%%%%%%%%%%%%%%%%%%%%%%%%%%%%%%%%%%%%%%%%%%%%%%%%%%%%%%%%%%%%%%%%%%%%%%%

%%%%%%%%%%%%%%%%%%%%%%%%%%%%%%%%%%%%%%%%%%%%%%%%%%%%%%%%%%%%%%%%%%%%%%%
\end{document}